\documentclass[graybox, secnum]{svmult}

\usepackage{mathptmx}       
\usepackage{helvet}         
\usepackage{courier}        
\usepackage{type1cm}        
\usepackage{makeidx}         
\usepackage{graphicx}        
\usepackage{multicol}        
\usepackage[bottom]{footmisc}
\usepackage{hyperref}        
\usepackage{soul}            
\hypersetup{colorlinks=true,urlcolor=blue}
\usepackage[square,numbers]{natbib}
\makeindex             

\usepackage{latexsym}
\usepackage{amssymb}
\usepackage{amsfonts}
\usepackage{amsmath}
\usepackage{graphicx} 
\usepackage{indentfirst}
\usepackage{bbm}
\usepackage{verbatim}
\usepackage{mathrsfs}
\usepackage{dsfont}
\usepackage{xcolor}
\usepackage{ytableau}

\newcommand {\cC}{{\cal C}}
\newcommand {\cD}{{\cal D}}
\newcommand {\cE}{{\cal E}}

\newcommand {\cH}{{\cal H}}

\newcommand {\cK}{{\cal K}}
\newcommand {\cL}{{\cal L}}
\newcommand {\cM}{{\cal M}}
\newcommand {\cN}{{\cal N}}

\newcommand {\cR}{{\cal R}}

\newcommand {\cT}{{\cal T}}

\newcommand {\cV}{{\cal V}}

\def\a{\alpha}

\def\b{\beta}
\def\c{\chi}
\def\d{\delta}
\def\e{\epsilon}
\def\f{\phi}
\def\g{\gamma}
\def\G{\Gamma}

\def\j{\psi}
\def\k{\kappa}
\def\l{\lambda}
\def\m{\mu}
\def\n{\nu}
\def\o{\omega}

\def\q{\theta}
\def\r{\rho}
\def\s{\sigma}
\def\t{\tau}

\def\x{\xi}

\def\D{\Delta}
\def\F{\Phi}
\def\J{\Psi}
\def\L{\Lambda}
\def\O{\Omega}
\def\P{\Pi}

\def\S{\Sigma}
\def\U{\Upsilon}
\def\X{\Xi}

\def\rd{{\rm d}}
\def\ri{{\rm i}}
\def\re{{\rm e}}

\newcommand{\ad}{{\dot{\alpha}}}                           
\newcommand{\bd}{{\dot{\beta}}}                            
\newcommand{\ve}{\varepsilon}                            
\newcommand{\cDB}{{\bar\cD}}                            

\renewcommand{\aa}{{\a\ad}}

\newcommand{\pa}{\partial}                           
\newcommand{\hf}{\frac12}

\newcommand{\vf}{\varphi}

\newcommand{\be}{\begin{equation}}
\newcommand{\ee}{\end{equation}}
\newcommand{\bea}{\begin{eqnarray}}
\newcommand{\eea}{\end{eqnarray}}
\newcommand{\non}{\nonumber}
\def\double #1{#1{\hbox{\kern-2pt $#1$}}}

\newcommand{\gd}{{\dot\g}}
\newcommand{\dd}{{\dot\d}}

\renewcommand{\ts}{{\tilde{\s}}}

\newif\ifdtup

\def\de{{\nabla}}                                         

\def\deb{{\bar{\de}}}

\newcommand{\bsubeq}{\begin{subequations}}
\newcommand{\esubeq}{\end{subequations}}

\newcommand{\sSL}{\mathsf{SL}}

\newcommand{\sSO}{\mathsf{SO}}
\newcommand{\sU}{\mathsf{U}}

\newcommand{\sOSp}{\mathsf{OSp}}

\newcommand{\scT}{{\mathscr{T}}}

\allowdisplaybreaks

\begin{document}
\title*{Superspace approaches to $\cN=1$ supergravity}
\author{S. M. Kuzenko,\thanks{corresponding author}
E. S. N. Raptakis
 and 
 G. Tartaglino-Mazzucchelli}
\institute{Sergei M. Kuzenko and Emmanouil S. N. Raptakis \at Department of Physics M013, The University of Western Australia 35 Stirling Highway, Perth W.A. 6009, Australia, \email{sergei.kuzenko@uwa.edu.au}, 
\email{emmanouil.raptakis@research.uwa.edu.au}
\and Gabriele Tartaglino-Mazzucchelli \at School of Mathematics and Physics, University of Queensland, St Lucia, Brisbane, Queensland 4072, Australia \email{g.tartaglino-mazzucchelli@uq.edu.au}}
%
%
\maketitle
\abstract{The superspace formalism for ${\cal N}=1$ supergravity  in four dimensions
is a powerful geometric setting to engineer off-shell supergravity-matter theories, including  higher-derivative couplings. This review provides a unified description of the three superspace approaches to ${\cal N}=1$ conformal supergravity: (i) conformal superspace; (ii) $\mathsf{U}(1)$ superspace; and (iii) the Grimm-Wess-Zumino formalism. The prepotential formulation for the latter is discussed. We briefly describe the known off-shell formulations for Poincar\'e and anti-de Sitter supergravity theories as conformal supergravity coupled to certain compensators. 
As simple applications of the formalism, we present the superfield equations of motion for various off-shell formulations for pure Poincar\'e and anti-de Sitter supergravity, and show that every solution of these equations is also a solution of the equations of motion for conformal supergravity.  
}

\section*{Keywords} 
Superconformal symmetry, Supergravity, Superspace
\vspace{0.5cm}
\begin{flushright}
{\it Dedicated to the creators of superfield supergravity}
\end{flushright}

\numberwithin{equation}{section}


\section{Introduction}\label{Section1}

Soon after the discovery of $\cN=1$ supergravity in four spacetime dimensions 
\cite{FvNF,DZ}
(and subsequent construction of the supersymmetric cosmological term \cite{Townsend})
 several off-shell formulations for this theory, 
 with different sets of auxiliary fields,  were developed. 
 These include the non-minimal \cite{Breitenlohner,Siegel77-80,SG}, 
old minimal \cite{Siegel77-77,WZ,old1,old2} and new minimal 
\cite{SohniusW1,SohniusW3} supergravity theories. 
The most general matter couplings are offered by the old minimal formulation 
\cite{GGRS,FGKV}. All interactions constructed in the framework of the new minimal as well as the non-minimal formulations are particular cases of those that can be realised within the old minimal theory \cite{FGKV}.

Traditionally, Refs. \cite{old1,old2} are credited with the discovery of old minimal supergravity, see e.g. \cite{Ferrara:1987ju}, since they have played a fundamental role in  the development of supergravity.
In fact, this off-shell theory was constructed for the first time using superfield techniques in an unpublished 1977 work by Siegel \cite{Siegel77-77} (which was difficult to digest at the time) and then re-discovered by Wess and Zumino \cite{WZ}, shortly before the publication of \cite{old1,old2}. It was explicitly shown \cite{SG,WZ78,WB}
that the component reduction of the superfield action for supergravity proposed in \cite{WZ} coincides with the component actions given in \cite{old1,old2}. It was also explicitly demonstrated \cite{SG,Siegel78} that the Wess-Zumino action \cite{WZ} is equivalent to the one proposed in \cite{Siegel77-77}. These are just simple examples of the power of superspace approaches to supergravity. 

Einstein's theory of gravity can be described using a Weyl invariant extension of the Einstein-Hilbert action by means of a compensating scalar field \cite{Deser,Zumino}.
In other words, ordinary gravity can be thought of as conformal gravity coupled to a compensator. As a generalisation of this idea, Poincar\'e supergravity can be realised as a locally superconformal invariant theory of supergravity \cite{KakuTownsend} in which the Weyl multiplet of conformal supergravity \cite{KTvN1, KTvN2} is coupled to a compensating scalar multiplet \cite{GGRS,FGKV}.\footnote{A similar idea was put forward earlier in \cite{Siegel77-77}.}
It turns out that all known off-shell formulations for Poincar\'e and anti-de Sitter (AdS) supergravity theories can be recast as conformal supergravity coupled to certain compensating multiplets. 
Different choices of a compensator correspond to different off-shell formulations for supergravity. This superconformal setting  has been developed in the conventional component approach \cite{Townsend:1979ki,Ferrara:1978rk,Kugo:1982cu,Kugo:1983mv} 
under the name ``superconformal tensor calculus''
 and 
has proved to be truly useful in order to formulate general two-derivative supergravity-matter systems 
and to study their dynamical properties, see \cite{FVP} for a review. In our opinion, it becomes especially powerful within superspace formulations for supergravity, which (i) provide remarkably compact expressions for general supergravity-matter actions; 
(ii) make manifest the geometric properties of such theories; and, most importantly, 
(iii) offer unique tools to generate higher-derivative couplings in matter-coupled supergravity. 

There are three fully fledged  approaches to describe $\cN=1$ conformal supergravity in superspace: (i) the Grimm-Wess-Zumino (GWZ) formalism \cite{GWZ} extending the Wess-Zumino formulation for on-shell supergravity \cite{WZ77}; (ii) the so-called $\sU(1)$ superspace proposed by Howe \cite{Howe}; and (iii) the $\cN=1$ conformal superspace approach developed by Butter \cite{ButterN=1}.
Conformal superspace  is an ultimate formulation for 
 conformal supergravity in the sense that any different off-shell formulation is either equivalent to it  or is obtained from it by partially fixing the gauge freedom. 
 In particular, $\sU(1) $ superspace can be obtained from a partial gauge fixing of the gauge group in conformal superspace.
 The $\cN=1$ superconformal tensor calculus reviewed in \cite{FVP,FT} is also a gauged fixed version of conformal superspace as demonstrated in \cite{ButterN=1}.
Recently, new supertwistor formulations were discovered for 
conformal supergravity theories  in diverse dimensions \cite{HL20}.
In the four-dimensional $\cN=1$ case, the supertwistor formulation is expected to be related to conformal superspace, however relevant technical details have not yet been worked out in the literature. 

Due to space limitations, in this review we are not able to discuss many important aspects of $\cN=1$ supergravity and its matter couplings. We apologise for the unavoidable omissions and missing references.

Our two-component spinor notation and conventions follow 
\cite{BK}, and are similar to those adopted in  \cite{WB}.  The only difference is that the spinor Lorentz generators $(\s_{ab})_\a{}^\b$ and 
$({\tilde \s}_{ab})^\ad{}_\bd$  used in \cite{BK} have an extra minus sign as compared with \cite{WB}, 
specifically $\s_{ab} = -\frac{1}{4} (\s_a \tilde{\s}_b - \s_b \tilde{\s}_a)$ and 
 $\tilde{\s}_{ab} = -\frac{1}{4} (\tilde{\s}_a {\s}_b - \tilde{\s}_b {\s}_a)$.


\section{Rigid and local superconformal transformations} 
\label{section2}

In this section we first review the structure of rigid superconformal transformations in Minkowski superspace ${\mathbb M}^{4|4}$. Then we introduce local superconformal transformations and describe the multiplet of conformal supergravity, following the approach due to Ogievetsky and Sokatchev \cite{OS}. It should be pointed out that the superconformal transformations in ${\mathbb M}^{4|4}$ were first studied by Sohnius \cite{Sohnius}. Our presentation follows \cite{BK}.

\subsection{Rigid superconformal transformations}
\label{section2.1}

We denote by $z^A =(x^a, \q^\a , \bar \q_\ad)$ the Cartesian coordinates
for Minkowski superspace ${\mathbb M}^{4|4}$,  
and use the notation $D_A =(\pa_a, D_\a , \bar D^\ad)$ 
for the  superspace covariant derivatives. 
The only non-trivial graded commutation relation is 
\bea
\{ D_\a , \bar D_\ad \}= 
-2\ri (\s^b)_{\a\ad}\pa_b
=-2\ri \pa_{\a\ad}~.
\eea

An infinitesimal superconformal  transformation 
$ z^A \to    z^A  + \d z^A$, with 
$\d z^A =\x  \,z^A = \Big(\x^a +\ri (\x\s^a \bar \q - \q \s^a \bar \x), \x^\a, \bar \x_\ad\Big), $ is generated by 
a {\it conformal Killing supervector field}
\bea
\x = {\overline \x} = \x^b  \pa_b + \x^\b D_\b
+ {\bar \x}_{\bd}  {\bar D}^{\bd}~.
\eea   
The defining property of $\x$ is that it takes every chiral superfield $\F$ to a chiral one, 
\bea
{\bar D}^{ \ad} \F =0 \quad \longrightarrow \quad 
{\bar D}^{\ad} (\x \F )=0~. 
\eea
This condition implies the relations 
\bea
{\bar D}^{ \ad }\x^\b =0~, \qquad
{\bar D}^{ \ad } \x^{ \bd \b} = 4{\rm i} \, \ve^{\ad \bd} \x^\b \quad \implies \quad
\x^\a = -\frac{\ri }{8} \bar D_{\ad } \x^{\ad \a }  
\eea
and their complex conjugates, 
and therefore
\bea
{\bar D}_{(\a } \x_{\b) \bd } =0~, \qquad {\bar D}_{(\ad } \x_{\b \bd )} =0 
\quad \implies \quad \pa_{(\a (\ad} \x_{\b) \bd)}=0~.
\eea
It follows that
\bea
[\x , D_\a ] = - (D_\a \x^\b) D_\b
= - K_\a{}^\b [\x] D_\b - 
\Big( \bar {\s} [\x] - \hf {\s} [\x] \Big) D_\a~.
\label{4Dmaster2} 
\eea
Here we have introduced chiral Lorentz  
($K_{\b\g}[\x] = K_{\g\b}[\x]$)
and super-Weyl ($\s[\x]$) parameters defined by 
\begin{subequations} 
\bea
K_{\a\b}[\x]&=& D_{(\a} \x_{\b)}~, \qquad \bar D_\gd K_{\a\b}[\x] =0~, \\
\s [\x] &= & \frac{1}{3} ( D_\a \x^\a + 2 \bar D^\ad \bar \x_\ad )
~, \qquad \bar D_\gd \s[\x] =0
~.
\eea
\end{subequations}
We recall that the Lorentz parameters with vector and spinor indices are related to each other as follows:
$K^{bc}[\x] = (\s^{bc})_{\b\g}K^{\b\g}[\x] - (\tilde{\s}^{bc})_{\bd\gd}\bar K^{\bd\gd}[\x] $.

The most general conformal Killing supervector field has the form 
\begin{subequations} \label{chiraltra}
\bea
 \x_+^{\ad\a} &=& a^{\ad\a}  +\hf (\s +{\bar \s})\, y^{\ad\a}
+{\bar K}^\ad{}_\bd  \,y^{\bd \a} +y^{\ad\b}K_\b{}^\a 
-y^{\ad \b} b_{\b \bd} y^{\bd \a} \non \\
&& \qquad +4{\rm i}\, {\bar \e}^\ad   \q^\a - 4 y^{\ad \b} \eta_\b \q^\a ~,
 \\
\x^\a &=& \e^\a + \big(\bar \s -\hf \s\big) \q^\a + \q^\b K_\b{}^\a 
-  \q^\b b_{\b\bd}   y^{\bd \a}
-{\rm i}\,{\bar \eta}_\bd y^{\bd \a} +2 \q^2 \eta^\a~,~~~~
\eea
\end{subequations}
where we have introduced the complex four-vector
\be
\x_+^a = \x^a + 2\ri \x \s^a  {\bar \q}~, 
\qquad \bar \x^a =\x^a~,
\ee
along with the complex bosonic coordinates $y^a = x^a +\ri \q \s^a \bar \q$ 
of the chiral subspace of ${\mathbb M}^{4|4}$. 
The constant bosonic parameters in \eqref{chiraltra}
correspond to the spacetime translation ($a^{\ad \a}$), 
 Lorentz transformation ($K_\b{}^\a,~{\bar K}^{\ad}{}_{\bd})$,
 special conformal transformation
($ b_{\a \bd}$), and  combined scale and $R$-symmetry transformations 
($\s =\t -\frac{2}{3} \ri \vf$). The constant fermionic parameters in \eqref{chiraltra}
correspond to the $Q$-supersymmetry ($\e^\a$) and $S$-supersymmetry 
($\eta_\a$) transformations. The constant parameters $K_{\a\b}$ and $\s$ are obtained 
from $K_{\a\b}[\x]$ and $\s[\x]$, respectively,  by setting $z^A=0$.

It is convenient to introduce a condensed notation for the superconformal parameters 
 \bea
 \l^{\tilde a} = (a^A, K^{ab} , \t, \vf , b_A)~, \quad 
 a^A:= (a^a, \e^\a , \bar \e_\ad)~, \quad b_A :=(b_a, \eta_\a , \bar \eta^\ad)~,
 \eea 
 as well as for the generators of the superconformal group
 \bea
 X_{\tilde a} = (P_A, M_{ab} ,{\mathbb D}, \mathbb{Y} , K^A)~, \quad 
 P_A:= (P_a, Q_\a , \bar Q^\ad)~, \quad K^A:=(K^a, S^\a , \bar S_\ad)~. 
 \label{generators}
 \eea 
The general conformal Killing supervector field on ${\mathbb C}^{4|2}$,
\bea
\x= \x^a_+ (y,\q) \frac{\pa}{\pa y^a} +\x^\a (y, \q)  \frac{\pa}{\pa \q^\a} 
\equiv \x^a_+ \partial/\partial y^a + \x^\a \pa_\a~,
\eea
may be written in the form:
\bea
\x = \l^{\tilde a} \x_{\tilde a} (X) = a^A \x_A (P) +\hf K^{ab} \x_{ab}(M) 
+\t \x({\mathbb D}) + \ri \vf \x(\mathbb{Y}) +b_A \x^A (K)~.~~
\eea
We read off the relevant supervector fields:
\begin{subequations}
\bea
\x_a(P) &=& \partial/\partial y^a ~, \quad \x_\a (P) = \pa_\a~, \quad 
\bar \x^\ad (P) = -2\ri (\tilde{\s}{}^c \q )^\ad \partial/\partial y^c ~, \\
\x_{ab}(M) &=& y_a \partial/\partial y^b - y_b \partial/\partial y^a +(\q\s_{ab} )^\g \pa_\g~,\\
\x({\mathbb D}) &=& y^c \partial/\partial y^c + \hf \q^\g \pa_\g~, \quad \x(\mathbb{Y}) = \q^\g \pa_\g~, \\
\x^a (K) &=& 2 y^a y^c \partial/\partial y^c -y^2 \partial/\partial y_a - (\q \s^a \tilde{\s}{}^c)^\g y_c \pa_\g~,\\
\x^\a (K) &=& 2 (\q\s^c \tilde{\s}{}^d)^\a y_d \partial/\partial y^c - 2\q^2 \ve^{\a\g} \pa_\g ~, \\
{\bar \x}_\ad (K) &=& \ri (\s^c)^\g{}_\ad y_c \pa_\g~. 
\eea
\end{subequations}

Making use of the above operators, we derive 
 the graded commutation relations for the superconformal algebra, 
 $\big[X_{\tilde a} , X_{\tilde b} \big\}= - f_{\tilde a \tilde b}{}^{\tilde c} X_{\tilde c}$,
 keeping in mind the relation
 \bea
 \x = \l^{\tilde a} \x_{\tilde a} (X)~ \to~ \d_\x =  \l^{\tilde a} X_{\tilde a} ~, \qquad 
 \big[\x_1 , \x_2 \big] ~\to ~- \big[\d_{\x_1} , \d_{\x_2}\big] ~.
 \eea
We start with the commutation relations for the conformal algebra:
\begin{subequations} \label{confal}
\begin{align}
&[M_{ab},M_{cd}]=2\eta_{c[a}M_{b]d}-2\eta_{d[a}M_{b]c}~, \phantom{inserting blank space inserting} \label{confal.16a}\\
&[M_{ab},P_c]=2\eta_{c[a}P_{b]}~, \qquad \qquad \qquad \qquad ~ [\mathbb{D},P_a]=P_a~,\\
&[M_{ab},K_c]=2\eta_{c[a}K_{b]}~, \qquad \qquad \qquad \qquad [\mathbb{D},K_a]=-K_a~,\\
&[K_a,P_b]=2\eta_{ab}\mathbb{D}+2M_{ab}~.
\end{align}
\end{subequations}
The $R$-symmetry generator $\mathbb Y$ commutes with all the generators of the conformal group. The superconformal algebra is obtained by extending the translation generator to $P_A$ and the special conformal generator to $K^A$. The commutation relations involving the $Q$-supersymmetry generators with the bosonic ones are:
 \begin{subequations} \label{superconfal1}
 \bea
 \big[M_{ab}, Q_\g \big] &=& (\s_{ab})_\g{}^\d Q_\d ~,\quad 
\big[M_{ab}, \bar Q^\gd \big] = (\tilde{\s}_{ab})^\gd{}_\dd \bar Q^\dd~,\\
\big[\mathbb{D}, Q_\a \big] &=& \hf Q_\a ~, \quad
\big[\mathbb{D}, \bar Q^\ad \big] = \hf \bar Q^\ad ~, \\
\big[\mathbb{Y}, Q_\a \big] &=&  Q_\a ~, \quad
\big[\mathbb{Y}, \bar Q^\ad \big] = - \bar Q^\ad ~,  \\
\big[K^a, Q_\b \big] &=& -\ri (\s^a)_\b{}^\bd \bar{S}_\bd ~, \quad 
\big[K^a, \bar{Q}^\bd \big] = 
-\ri ({\s}^a)^\bd{}_\b S^\b ~.
 \eea
 \end{subequations}
The commutation relations involving the $S$-supersymmetry generators 
with the bosonic operators are: 
\begin{subequations} \label{superconfal2}
 \bea
\big [M_{ab} , S^\g \big] &=& - (\s_{ab})_\b{}^\g S^\b ~, \quad
\big[M_{ab} , \bar S_\gd \big] = - (\ts_{ab})^\bd{}_\gd \bar S_\bd~, \\
\big[\mathbb{D}, S^\a \big] &=& -\hf S^\a ~, \quad
\big[\mathbb{D}, \bar S_\ad \big] = -\hf \bar S_\ad ~, \\
\big[\mathbb{Y}, S^\a \big] &=&  -S^\a ~, \quad
\big[\mathbb{Y}, \bar S_\ad \big] =  \bar S_\ad ~,  \\
\big[ S^\a , P_b \big] &=& \ri (\s_b)^\a{}_\bd \bar{Q}^\bd ~, \quad 
\big[\bar{S}_\ad , P_b \big] = 
\ri ({\s}_b)_\ad{}^\b Q_\b ~.
 \eea
 \end{subequations}
Finally, the anti-commutation relations of the fermionic generators are: 
\begin{subequations}\label{superconfal3}
\bea
\{Q_\a , \bar{Q}^\ad \} &=& - 2 \ri  (\s^b)_\a{}^\ad P_b=- 2 \ri   P_\a{}^\ad~, \\
\{ S^\a , \bar{S}_\ad \} &=& 2 \ri  (\s^b)^\a{}_\ad K_b=2 \ri   K^\a{}_\ad
~, 
\label{superconfal3-2}\\
\{ S^\a , Q_\b \} &=& 2 \d^\a_\b \mathbb{D} - 4  M^\a{}_\b 
- 3 \d^\a_\b \mathbb{Y} ~, \\
\{ \bar{S}_\ad , \bar{Q}^\bd \} &=& 2  \d^\bd_\ad \mathbb{D} 
+ 4  \bar{M}_\ad{}^\bd +  3\d_\ad^\bd \mathbb{Y}  ~,
\eea
\end{subequations}
where $M_{\a\b}=\hf(\s^{ab})_{\a\b}M_{ab}$ and $\bar{M}_{\ad\bd}=-\hf(\ts^{ab})_{\ad\bd}M_{ab}$. Note that all remaining (anti-)commutators not explicitly listed above vanish identically.

The graded commutation relations
\eqref{confal} -- \eqref{superconfal3} constitute
the $\cN=1$ superconformal algebra, ${\mathfrak{su}}(2,2|1)$. Its generators obey
the graded Jacobi identity
\bea
(-1)^{\varepsilon_{\tilde{a}}  \varepsilon_{\tilde{c}}}[X_{\tilde{a}}, [X_{\tilde{b}}, X_{\tilde{c}} \} \} 
~+~ \text{(two cycles)}
= 0 \ ,
\label{Jacobi-0}
\eea
where $\varepsilon_{\tilde{a}} = \varepsilon(X_{\tilde{a}})$ is the Grassmann parity of the generator $X_{\tilde{a}}$. Making use of 
$\big[X_{\tilde a} , X_{\tilde b} \big\}= - f_{\tilde a \tilde b}{}^{\tilde c} X_{\tilde c}$,
the Jacobi identities are equivalently written as
\bea 
f_{[\tilde{a}\tilde{b}}{}^{\tilde{d}} f_{|\tilde{d}| \tilde{c} \} }{}^{\tilde{e}} = 0 \ .
\eea

It remains to discuss superconformal transformation laws for superfields. Here we restrict our discussion to {\it primary superfields}. Given a conformal Killing supervector field $\x$, the corresponding infinitesimal superconformal transformation acts on a primary tensor superfield $U$ (with suppressed indices) by the rule 
\bea
\d_\x U =  \cK[\x] U~,
\qquad \cK[\x] = \x + \hf K^{ab}[\x] M_{ab}  +  p \s[\x] + q \bar \s[\x] ~.
\label{primaryTL}
\eea
Here the parameters $p$ and $q$ are related to the dimension (Weyl weight) $w$ and $\sU(1)_R$ charge $c$ of $U$ as follows: $w = p+q$ and $p-q = - \frac 32 c$.
The Lorentz generators $M_{ab}$ in \eqref{primaryTL} act on the indices of $U$.
The commutation relation for these matrices differs by overall sign from \eqref{confal.16a}. This is due to the fact that, in conformal (super)gravity, 
subsequent transformations are applied as follows: $\d_{\x_2} \d_{\x_1} U = 
\d_{\x_2}  \cK[\x_1] U =  \cK[\x_1] \d_{\x_2} U =  \cK[\x_1]  \cK[\x_2] U$, see \cite{FVP} for  more details.


\subsection{Superconformal transformations and complex geometry}

Minkowski superspace ${\mathbb M}^{4|4}$ is embedded in the so-called chiral  
superspace ${\mathbb C}^{4|2}$, parametrised by complex coordinates
$y^a$ and $\q^\a$, 
as the real surface 
\bea
\hf ( y^a - \bar y^a ) = \ri \q \s^a \bar \q ~, \qquad  \hf( y^a +\bar y^a) =x^a
~.
\label{MinkowskiSS}
\eea
This  is a special member of a family of real superspaces
$\cM^{4|4} (\cH)$  embedded in ${\mathbb C}^{4|2}$ by the rule
\bea
\hf ( y^a - \bar y^a)  = \ri  \cH^a (x, \q, \bar \q) ~, \qquad  \hf( y^a +\bar y^a) =x^a~,
\label{embedding223}
\eea
where the four real bosonic functions $ \cH^a (x, \q, \bar \q) $ 
may be arbitrary. With respect to 
the super-Poincar\'e transformations 
\bea
\d y^a = a^a - K^a{}_b y^b +2\ri \q \s^a \bar \e~, \qquad 
 \d \q^\a = \e^\a + \hf K^{bc} (\q\s_{bc})^\a  ~,
\label{super-Poincare}
\eea
$ \cH^a(z) - \q \s^a \bar \q$ proves to be a vector superfield.
What is special about Minkowski superspace, $\cM^{4|4} (\q \s \bar \q)$, is the fact that \eqref{MinkowskiSS} is the unique 
surface of the type $\cM^{4|4} (\cH)$  which is invariant under the  super-Poincar\'e transformations \eqref{super-Poincare}.

It turns out that the superconformal transformations  \eqref{chiraltra} are the most general holomorphic transformations 
on ${\mathbb C}^{4|2}$ 
of the form
\bea
\d y^a = \l^a (y, \q) ~, \quad \d \q^\a = \l^\a (y, \q)~, 
\label{holom}
\eea
which leave invariant the superspace ${\mathbb M}^{4|4}$ defined by \eqref{MinkowskiSS}. This remarkable result indicates that (i) arbitrary holomorphic transformations \eqref{holom} should be interpreted as local superconformal ones; 
and (ii) $\cH^a (x, \q, \bar \q) $ should be used to describe conformal supergravity,  a supersymmetric extension of conformal gravity.


\subsection{Local superconformal transformations} \label{section2.3}

Following \cite{OS}, the gauge group of conformal supergravity is postulated to be the 
supergroup of holomorphic reparametrisations of ${\mathbb C}^{4|2} $
\bea
y^m \to y'^m=f^{m}  (y, \q) ~, \quad \q^\m \to \q'^\m = f^\m (y,\q) ~, \quad 
{\rm Ber} \left( \frac{\pa(y', \q') }{\pa(y,\q)}\right) \neq 0~.~~
\eea
In curved superspace, we distinguish between curved and flat-space indices.
Latin and Greek letters from the middle of each alphabet are used for curved-space indices. Letters from the beginning of each alphabet denote flat-space indices.

In practise, it suffices to work with infinitesimal holomorphic transformations, 
\bea
y^m \to y'^m = y^m - \l^m(y,\q) ~, \quad \q^\m \to \q'^\m = \q^\m -\l^\m(y,\q).
\eea
When restricted to $\cM^{4|4} (\cH)$, this transformation acts as follows
\begin{subequations}
\bea
x^m & \to & x'^m = x^m - \hf \l^m (x +\ri \cH, \q) - \hf \bar \l^m (x-\ri \cH, \bar \q) ~,\\
\q^\m &\to & \q'^\m = \q^\m - \l^\m (x+\ri \cH, \q) ~,
\eea
as well as 
\bea
\cH'^m (x', \q' , \bar \q') = \cH^m(x,\q ,\bar \q) 
+\frac{\ri}{2} \l^m (x +\ri \cH, \q) - 
\frac{\ri}{2} 
\bar \l^m (x-\ri \cH, \bar \q) ~.~~
\eea
\end{subequations}
From here it follows that $\d \cH^m = \cH'^m (x, \q , \bar \q) - \cH^m(x,\q ,\bar \q) $ is given by
\bea
\d \cH^m  &=& \frac{\ri}{2} ( \l^m - \bar \l^m) 
+\Big( \hf (\l^n +\bar \l^n) \pa_n +\l^\m \pa_\m +\bar \l_{\dot \m}\pa^{\dot \m} \Big) \cH^m ~, \non \\
\l^m &=&  \l^m (x +\ri \cH, \q) ~,\quad \l^\m=\l^\m (x+\ri \cH, \q) ~.
\label{H-transformation}
\eea
This is the gauge transformation law of $\cH^m$.

Making use of the gauge freedom for $\cH^m$ allows one to choose a gauge condition 
\bea
 \cH^m (x,\q,\bar \q) &=& \q \s^a\bar \q e_a{}^m (x) - \ri \bar \q^2 \q^\a \J^m_\a (x)
 +\ri \q^2 \bar \q_\ad \bar \J^{m \ad} (x) \non \\
&& + \q^2 \bar \q^2 \Big(A^m (x) - \frac 14 e_a{}^m (x) \ve^{abcd} \o_{bcd}(x) \Big)
~.~~~
\label{WZgauge}
 \eea
 Here $\o_{abc}=-\o_{acb} = e_a{}^m \o_{mbc}$ is the torsion-free Lorentz connection associated with the vielbein $e^a =\rd x^m e_m{}^a $ and its dual frame field $e_a = e_a{}^m \pa_m $:
\bea
 \o_{abc} =- \hf \left( \cC_{bca} + \cC_{acb} -\cC_{abc} \right)~,\qquad 
 \big[ e_a , e_b \big] =\cC_{ab}{}^c e_c~.
 \label{T-Fconnection}
 \eea
 The residual gauge freedom, which preserves the condition \eqref{WZgauge}, is given by  
 \begin{subequations} \label{WZparameters}
 \bea
  \l^m ( \q) &=& {\mathfrak a}^m  +2\ri \q \s^a \bar \e   e_a{}^m 
 -2\q^2 \bar \e \bar \J^m  ~, \qquad \bar {\mathfrak a}^m = {\mathfrak a}^m ~, \\
 \l^\a (\q) &=& \e^\a + \q^\a \Big( \hf \t + \ri \vf \Big) +\q^\b K_\b{}^\a  \non \\
 &&  +\q^2 \Big[\eta^\a 
  - \hf(\bar \e \tilde{\s}^a)^\a
 \Big( \ri \o^b{}_{ba}  +\frac 12 \ve^{abcd} \o_{bcd} \Big) \Big]~,
 \eea
 \end{subequations}
 with 
  $K_\a{}^\b = \hf K^{ab} (\s_{ab})_\a{}^\b$.
  Keeping in mind the structure of conformal Killing supervector fields \eqref{chiraltra},
 we can give the following interpretations to the gauge parameters \eqref{WZparameters}.
 The bosonic parameters 
 correspond to the general coordinate (${\mathfrak a}^m $), local Lorentz 
 ($K_\a{}^\b $), Weyl ($\tau$) and local chiral 
 ($\vf$) transformations. The fermionic parameters correspond to the local 
 $Q$-supersymmetry $(\e^\a$) and $S$-supersymmetry ($\eta^\a$) transformations. 
 
The superfield gauge transformation \eqref{H-transformation} allows us to work out the transformation laws of the component fields in \eqref{WZgauge}. Choosing 
${\mathfrak a}^m \neq0$ and switching off the other parameters in \eqref{WZparameters} gives 
 \bea
\d_{\mathfrak a} e_a = \big[ {\mathfrak a}, e_a\big ]
~,\quad 
\d_{\mathfrak a} \J_\a = \big[ {\mathfrak a}, \J_\a \big ]
~, \quad 
\d_{\mathfrak a} A = \big[ {\mathfrak a}, A \big ]~,
\label{GCT}
\eea 
 where we have introduced the first-order operators 
 ${\mathfrak a} = {\mathfrak a}^m\pa_m$, $\J_\a = \J^m_\a \pa_m$ and
 $A= A^m \pa_m$. Next, choosing $K_\a{}^\b 
  \neq 0$
  and switching off the other parameters in \eqref{WZparameters} gives 
\bea
\d_K e_a = K_a{}^b e_b~, \quad \d_K \J_\a = K_\a{}^\b \J_\b ~, 
\quad \d_K A=0~.
\label{localL}
\eea
 The transformation laws \eqref{GCT} and \eqref{localL}
 allow us to interpret the field $e_a{}^m$ as the inverse vielbein. They also show that $\J^m_\a$ transforms as a world vector and a Weyl spinor, 
 while $A^m$ is a vector field. Next, choosing $\hf \t +\ri \vf \neq 0$
  and switching off the other parameters in \eqref{WZparameters} gives the Weyl ($\t$) and local chiral ($\vf$) transformations
\bea
\d_\t e_a{}^m &=& \t e_a{}^m~, \quad \d_\t \J^m_\a = \frac 32 \t \J^m_\a~, \quad 
\d_\t A_m =0~; \\
\d_\vf e_a{}^m &=& 0 ~, \quad \d_\vf \J^m_\a = -\ri \vf \J^m_\a~, \quad 
\d_\vf A_m = \pa_m \vf~.
\eea
Here we have introduced the one-form $A_m =  g_{mn} A^n$, 
where $g_{mn} (x)= e_m{}^a (x)e_n{}^b (x)\eta_{ab} $ is the Lorentzian metric associated with the vielbein $e_m{}^a$. It follows that $A_m$ is the gauge field for the chiral $\sU(1)_R$ group. Since $\cH^m$ contains the inverse vielbein at the component level, it is called the {\it gravitational superfield}. 

It remains to consider local supersymmetry transformations. Choosing $\eta^\a \neq 0$ 
and switching off the other parameters in \eqref{WZparameters} gives the $S$-supersymmetry transformation laws
\bea
\d_{\eta} e_a{}^m=0~, \quad \d_{\eta} \J^m_\a = e_a{}^m(\s^a \bar \eta)_\a ~,
\quad \d_{\eta} A^m = \ri \big(\bar \eta \bar \J^m - \eta \J^m \big)~. 
\label{S-SUSY}
\eea
Finally, for the $Q$-supersymmetry transformation we obtain 
\begin{subequations} \label{Q-SUSY}
\bea
\d_\e e_a{}^m &=& \ri \big( \J^m \s_a \bar \e - \e \s_a \bar \J^m  \big) ~, \\
\d_\e \J^m_\a &=& - (\s^a \tilde{\s}^b \nabla_a \e )_\a e_b {}^m + 2\ri A^m \e_\a~, \\
\d_\e A^m &=& - \frac{\ri}{4} e_a{}^m \ve^{abcd} \nabla_b 
\big( \e \s_c \bar \J^n - \J^n \s_c \bar \e \big) e_{n d} 
- \hf \nabla_a \big( \e \s^a  \bar \J^m + \J^m \s^a \bar \e\big) \non \\
&& +  \big( \nabla_n \e \s^a \bar \J^n + \J^n \s^a \nabla_n \bar \e \big) e_a{}^m
~.
\eea
\end{subequations}
Here $\nabla_n $ and $\nabla_a = e_a{}^n \nabla_n$ are standard torsion-free covariant derivatives, 
in particular 
\bea
\nabla_n e_a{}^m = 0~, \qquad  
\nabla_n \J^m_\a = \pa_n \J^m_\a - \hf \o_n{}^{bc} (\s_{bc}  )_\a{}^\b \J_\b^m 
+  \G^m_{n r} \J_\a^r ~,
\eea
where the Lorentz connection is given by \eqref{T-Fconnection}  and $\G^m_{nr} $ denotes the Christoffel symbols.

The $S$ and $Q$-supersymmetry transformation laws can be rewritten in a more convenient and familiar form if the dynamical fields $e_a{}^m$, $\J^m_\a$ and $A$ are replaced with $e_m{}^a$, $\J_{m\a} = g_{mn} \J^n_\a$ and 
\be
{\mathfrak A}_m = A_m - \hf \J^n \s_m \bar \J_n 
+\hf \big( \J_m \s_n \bar \J^n + \J^n \s_n \bar \J_m \big)
+\frac{\ri}{8} e_m{}^a \ve_{abcd} \J^b \s^c \bar \J^d ~.
\ee
 Then the $S$-supersymmetry transformation turns into 
 \bea
\d_{\eta} e_m{}^a=0~, \quad \d_{\eta} \J_{m\a} = (\s_m \bar \eta)_\a ~,
\quad \d_{\eta} {\mathfrak A}_m = \frac 34  \ri \big( \eta \J_m - \bar \eta \bar \J_m \big) ~. 
\label{S-SUSY2}
\eea
It turns out that the simplest version of the $Q$-supersymmetry transformation corresponds to the variation 
\begin{subequations}
\bea
\hat \d_\e &=& \d_\e + \d_{\eta(\e) } + \d_{K(\e)} + \d_{\vf(\e)} ~, \\
\eta^\a (\ve) &=& - \ri (\nabla_b \bar \e \tilde{\s}^b)^\a -\e^\a \bar \J^n \bar \J_n 
+ \hf (\bar \e \tilde{\s}_a)^\a\J^n \s^a \bar \J_n \non \\
&& -\frac{\ri}{4} (\bar \e \tilde{\s}_a)^\a \ve^{abcd} \J_b \s_c \bar \J_d~, \\
K_{\a\b}(\e)&=& \frac{\ri}{2} \Big[ (\s_n \bar \J^n)_\a \e_\b + (\s_n \bar \J^n)_\b \e_\a 
- (\s_n \bar \e)_\a \J^n_\b - (\s_n \bar \e)_\b \J^n_\a \Big]~,~\\
\vf(\e) &=&  \hf \big( \J^n \s_n \bar \e + \e \s_n \bar \J^n \big) ~.
\eea
\end{subequations} 
Then we end up with the following transformation laws 
\begin{subequations}
\label{2.43}
\bea
\hat \d_\e e_m{}^a &=& \ri \big( \e \s^a \bar \J_m - \J_m \s^a \bar \e \big) ~,\\
\hat \d_\e \J_{m } &=& 2 \hat \nabla_m \e
= 2(\nabla_m - \hf \hat{\o}_m{}^{bc} \s_{bc} + \ri  {\mathfrak A}_m \big) \e ~,\\
\hat \d_\e {\mathfrak A}_m &=& 
\hf \e\s^n \big( \hat \nabla_m \bar \J_n -
\hat \nabla_n \bar \J_m \big)
- \hf \big( \hat \nabla_m \J_n -\hat \nabla_n \J_m \big) \s^n \bar \e \non \\
&&-\frac{\ri}{4} g_{mn}\ve^{nijk} 
\big( \hat \nabla_i \J_j \s_k \bar \e - \e \s_k \hat \nabla_i \bar \J_j \big)~,
\eea
\end{subequations}
where we have introduced the 
covariant derivative with torsion
\begin{subequations}
\label{2.44}
\bea
\hat \nabla_m &=& \nabla_m - \hf \hat{\o}_m{}^{bc} M_{bc} - \ri w {\mathfrak A}_m ~,\\
\hat{\o}_{cab} &=& -\hf \big( \hat C_{abc} + \hat C_{acb} - \hat C_{bca} \big) ~, \quad
 \hat C_{abc} = \frac{\ri}{2} \big( \J_a \s_c \bar \J_b - \J_b \s_c \bar \J_a \big)~,~~~
\eea
\end{subequations}
with $w$ being the $\sU(1)_R$ charge of a field $\U$ with the $\sU(1)_R$ transformation law $\U \to  \re^{\ri w \vf} \U$.

It follows from the above analysis that the gauge fields 
$\big\{ e_m{}^a , \J_{m\a} , \bar \J_{m}^\ad , {\mathfrak A}_m \big\}$ form a multiplet under 
the local  $S$ and $Q$-supersymmetry transformations. It will be referred to as the {\it reduced Weyl multiplet}\footnote{The Weyl multiplet also includes a dilatation gauge field, $b_m$, but this proves to describe purely gauge degrees of freedom, see e.g.
\cite{FVP,FT} for reviews and subsection \ref{section3.4} below.} of conformal supergravity
since it is a supersymmetric generalisation of conformal gravity in which
the gauge group includes the Weyl transformations $e_m{}^a (x) \to  \re^{-\t (x) }e_m{}^a (x)$.


\section{Conformal superspace}\label{N1_conformal-superspace}

In the previous section we have reviewed a simple approach to obtain 
the reduced Weyl multiplet of conformal supergravity from superspace. In this setting, conformal supergravity is described by the gravitational superfield $\cH^m$,  which defines the embedding of curved superspace 
$\cM^{4|4}(\cH)$ in the chiral superspace ${\mathbb C}^{4|2}$, eq. \eqref{embedding223}. Although this approach is elegant and geometric, it is not covariant and does not offer powerful tools to construct manifestly gauge-invariant supergravity actions and to engineer general couplings of supergravity to matter. 
Such tools are provided by the so-called conformal superspace approach  \cite{ButterN=1}, which is reviewed in the present section.


\subsection{Gauging the superconformal algebra in superspace} \label{Gauging}

Conformal superspace is a gauge theory of the superconformal algebra. It can be identified with a pair $(\cM^{4|4}, \nabla)$. Here $\mathcal{M}^{4|4}$ denotes a supermanifold parametrised by local  coordinates 
$z^M = (x^m, \q^\m, \bar \q_{\dot \m})$, and $\nabla$ is a covariant derivative associated with the superconformal algebra. We recall that the generators $ X_{\tilde a}$ of the superconformal algebra are given by eq. \eqref{generators}. They can be grouped in two disjoint subsets,
 \bea 
 X_{\tilde a} = (P_A, X_{\underline{a}} )~, \qquad
 X_{\underline{a}} =
( M_{ab} ,{\mathbb D}, \mathbb{Y} , K^A)~,
 \eea 
 each of which constitutes a superalgebra:
\bsubeq\label{RigidAlgebra}
\begin{align}
[P_{ {A}} , P_{ {B}} \} &= -f_{{ {A}} { {B}}}{}^{{ {C}}} P_{ {C}}
	\ , \\
[X_{\underline{a}} , X_{\underline{b}} \} &= -f_{\underline{a} \underline{b}}{}^{\underline{c}} X_{\underline{c}} \ , \\
[X_{\underline{a}} , P_{{B}} \} &= -f_{\underline{a} { {B}}}{}^{\underline{c}} X_{\underline{c}}
	- f_{\underline{a} { {B}}}{}^{ {C}} P_{ {C}}
	\ . \label{mixing}
\end{align}
\esubeq
Here the structure constants $f_{{ {A}}{ {B}}}{}^{ {C}}$ contain only one non-zero component, which is $f_{\a}{}^\bd{}^c = 2 \ri \,(\s^c)_\a{}^\bd$.

In order to define the covariant derivatives, $\nabla_A= (\nabla_a, \nabla_\a, \bar \nabla^\ad)$, we associate with each generator $X_{\underline{a}} =
( M_{ab} ,{\mathbb D}, \mathbb{Y} , K^A) = ( M_{ab} ,{\mathbb D}, \mathbb{Y} ,K^a, S^\a , \bar S_\ad)$
a connection one-form 
$\omega^{\underline{a}} = (\O^{ab},B,\Phi,\frak{F}_{A})= (\O^{ab},B,\Phi,\frak{F}_{a},\frak{F}_{\a},
\bar{\frak{F}}^{\ad})= \rd z^M \omega_M{}^{\underline{a}}$,  
and with $P_{ {A}}$ a supervielbein one-form
$E^{ {A}} = (E^a, E^\a,\bar{E}_\ad) = \rd z^{ {M}} E_M{}^A$ (the latter will be often referred to as the vielbein). 
It is assumed that the supermatrix $E_M{}^A$ is nonsingular, $E:= {\rm Ber} (E_M{}^A) \neq 0$, 
and hence there exists a unique inverse supervielbein. The latter is given by 
the supervector fields $E_A = E_A{}^M (z)\pa_M $, with 
$\pa_M = \pa /\pa z^M$, which constitute a new basis for the tangent space at each
  point $z^M \in \cM^{4|4}$. The supermatrices $E_A{}^M $ and $E_M{}^A$ satisfy the 
  properties $E_A{}^ME_M{}^B=\d_A{}^B$ and $E_M{}^AE_A{}^N=\d_M{}^N$.
With respect to  the basis $E^A$,  the connection is expressed as 
$\omega^{\underline{a}} =E^B\omega_B{}^{\underline{a}}$, 
where $\omega_B{}^{\underline{a}}=E_B{}^M\omega_M{}^{\underline{a}}$. 
 The {\it covariant derivatives} are given by 
\bea\label{nablaA}
\nabla_A 
&=& E_A  - \o_A{}^{\underline b} X_{\underline b}=
E_A -  \hf \Omega_A{}^{bc} M_{bc} - B_A \mathbb{D} - \ri \Phi_A \mathbb{Y}
- \mathfrak{F}_{AB} K^B~.~~~
\eea
They can be recast in terms of one-forms 
\be\label{eq:covD}
\nabla = \rd - \omega^{\underline{a}} X_{\underline{a}} \ , \quad \nabla = E^A \nabla_A \ .
\ee
The translation generators $P_B$ do not show up in \eqref{nablaA} and 
\eqref{eq:covD}. It is assumed that the operators $\nabla_A$ replace $P_A$ and obey the graded commutation relations
\be
[ X_{\underline{b}} , \nabla_A \} = -f_{\underline{b} A}{}^C \nabla_C
	- f_{\underline{b} A}{}^{\underline{c}} X_{\underline{c}} \ ,
\ee
compare with \eqref{mixing}.
In particular, the algebra of $K^A$ with $\nabla_B$ is given by
\begin{subequations}\label{Knabla}
\begin{align}
[K^a, \nabla_b] &= 2 \delta^a_b  \mathbb{D} + 2 M^{a}{}_b
 ~, \label{Knabla.a}\\
\{ S^\a , \nabla_\b \} &=  \d^\a_\b \big(2\mathbb{D} - 3  \mathbb{Y} \big)- 4  M^\a{}_\b  
~,\label{Knabla.b}\\
\{ \bar{S}_\ad , \bar{\nabla}^\bd \} &=   \d^\bd_\ad \big(2 \mathbb{D} +  3 \mathbb{Y} \big) 
+ 4  \bar{M}_\ad{}^\bd 
~, \label{Knabla.c} \\
[K^a, \nabla_\b] &= -\ri (\s^a)_\b{}^\bd \bar{S}_\bd \ , \qquad \qquad \qquad[K^a, \bar{\nabla}^\bd] = 
-\ri ({\s}^a)^\bd{}_\b S^\b ~, \label{Knabla.d} \\
[S^\a , \nabla_b] &= \ri (\s_b)^\a{}_\bd \bar{\nabla}^\bd \ , \qquad \qquad \quad \qquad[\bar{S}_\ad , \nabla_b] = 
\ri ({\s}_b)_\ad{}^\b \nabla_\b \ , \label{Knabla.e}
\end{align}
\end{subequations}
where all other graded commutators vanish.

By definition, the gauge group of conformal supergravity  is generated 
by local transformations of the form
\begin{subequations}\label{SUGRAtransmations}
\bea
\delta_\cK \nabla_A &=& [\cK,\nabla_A] \ , \\
\cK &=& \xi^B \nabla_B +  \L^{\underline{b}} X_{\underline{b}}
=  \xi^B \nabla_B+ \hf K^{bc} M_{bc} + \S \mathbb{D} + \ri \rho \mathbb{Y}
+ \L_B K^B \ ,
\eea
\end{subequations}
where  the gauge parameters satisfy natural reality conditions. 
In applying
eq. \eqref{SUGRAtransmations}, 
we interpret that
\begin{subequations}
\bea
 \nabla_A \xi^B &:=& E_A \xi^B + \omega_A{}^{\underline{c}} \xi^D f_{D \underline{c}}{}^{B} \ , \\
\nabla_A \L^{\underline{b}} &:=& E_A \L^{\underline{b}}
+ \omega_A{}^{\underline{c}} \xi^D f_{D \underline{c}}{}^{\underline{b}}
+ \omega_A{}^{\underline{c}} \L^{\underline{d}} f_{\underline{d}\underline{c}}{}^{\underline{b}} \ .
\eea
\end{subequations}
Then it follows from \eqref{SUGRAtransmations} that 
\begin{subequations}\label{3.8}
\begin{align}
\delta_\cK E^A &= \rd \xi^A + E^B \L^{\underline c} f_{\underline c B}{}^A
	+ \omega^{\underline b} \xi^{C} f_{C \underline b}{}^{A}
	+ E^B \xi^{C} \scT_{C B}{}^A~, \\
\delta_\cK \omega^{\underline a} &= \rd \L^{\underline a}
	+ \omega^{\underline b} \L^{\underline c} f_{\underline c \underline b}{}^{\underline a}
	+ \omega^{\underline b} \xi^{C} f_{C \underline b}{}^{\underline a}
	+ E^B \L^{\underline c} f_{\underline c B}{}^{\underline a}
	+ E^B \xi^{C}  \mathscr{R}_{C B}{}^{\underline a}~.
\end{align}
\end{subequations}
Here we have made use of the graded commutation relations
\bea
\label{3.9}
[ \nabla_A , \nabla_B \} = - \scT_{AB}{}^C \nabla_C - \mathscr{R}_{AB}{}^{\underline{c}} X_{\underline{c}} 
\ ,
\eea
where $ \scT_{AB}{}^C $ and $ \mathscr{R}_{AB}{}^{\underline{c}} $ denote the torsion and the curvature, respectively. They can be recast in terms of two-forms
\begin{subequations} \label{TRexpComp}
\begin{align}
\scT^A &:= \hf E^C \wedge E^B \scT_{BC}{}^A = \rd E^A - E^C \wedge \omega^{\underline{b}} \,f_{\underline{b} C}{}^A \ , \\
\mathscr{R}^{\underline{a}} &:= \hf E^C \wedge E^B \mathscr{R}_{BC}{}^{\underline{a}} = \rd \omega^{\underline{a}}
	- E^C \wedge \omega^{\underline{b}} \, f_{\underline{b} C}{}^{\underline{a}}
	- \hf \omega^{\underline{c}} \wedge \omega^{\underline{b}} \,
		f_{\underline{b} \underline{c}}{}^{\underline{a}} \ .
\end{align}
\end{subequations}
Making use of the graded Jacobi identity
\be 
0 = (-1)^{\ve_{\underline{a}} \ve_C } 
[ X_{\underline{a}} , [ \nabla_B , \nabla_C \} \} ~+~\text{(two cycles)}
\ee
we derive the action of $X_{\underline{a}} $ on the geometric objects
\begin{subequations} \label{313}
\begin{align}
X_{\underline{a}} \scT_{BC}{}^D =&
- (-1)^{\varepsilon_{\underline{a}} (\varepsilon_{B}+\varepsilon_{C})} 
\scT_{BC}{}^{ {E}} f_{ {E} \underline{a}}{}^D
- 2 f_{\underline{a} [B}{}^{ {E}} \scT_{| {E}| C\}}{}^D
- 2 f_{\underline{a} [B}{}^{\underline{e}} f_{|\underline{e}| C\}}{}^D \ , \\
X_{\underline{a}} \mathscr{R}_{BC}{}^{\underline{d}} =&
- (-1)^{\varepsilon_{\underline{a}} (\varepsilon_{B}+\varepsilon_{C})} 
\Big(\scT_{BC}{}^{ {E}} f_{ {E} \underline{a}}{}^{\underline{d}}
+ \mathscr{R}_{BC}{}^{\underline{e}} f_{\underline{e}\underline{a} }{}^{\underline{d}}\Big)
- 2 f_{\underline{a} [B}{}^{ {E}} \mathscr{R}_{| {E}| C \}}{}^{\underline{d}} \non\\
&- 2 f_{\underline{a} [B}{}^{\underline{e}} f_{|\underline{e}| C \}}{}^{\underline{d}} \ .
\end{align}
\end{subequations}

The supergravity gauge group acts on a conformal tensor superfield 
$U$ (with suppressed indices) as 
\bea 
\label{3.14}
\d_{\cK} U = \cK U \ .
\eea
The torsion $ \scT_{AB}{}^C $ and the curvature $ \mathscr{R}_{AB}{}^{\underline{c}} $ 
are conformal tensor superfields, for which the action of the generators $X_{\underline{a}} $ is specified by the relations \eqref{313}.
Of special significance are primary superfields. 
A tensor superfield $U$ (with suppressed indices) is said to be \emph{primary} if it is characterised by the properties 
\bea
K^A U = 0~, \quad \mathbb D U = w U~,  \quad {\mathbb Y} U = c U~,
\label{PrimaryDimensionCharge}
\eea
for some real constants $w$ and $c$ which are called the dimension (or Weyl weight)  and $\sU(1)_R$ charge of $U$, respectively.
It follows from  \eqref{superconfal3-2} that if a superfield is annihilated by the $S$-supersymmetry generators, then it is necessarily primary.

Let us summarise some important features of the gauging procedure. In curved superspace, the 
superconformal algebra \eqref{RigidAlgebra} is replaced with 
\begin{subequations}\label{gaugedSCA}
\begin{align}
[X_{\underline{a}} , X_{\underline{b}} \} &
= -f_{\underline{a} \underline{b}}{}^{\underline{c}} X_{\underline{c}} \ ,
\label{eq:XwithX} \\
[X_{\underline{a}} , \nabla_B \} &= - f_{\underline{a} B}{}^C \nabla_C -f_{\underline{a} B}{}^{\underline{c}} X_{\underline{c}} \ , \label{eq:XwithNabla} \\
[\nabla_A , \nabla_B \} &= -\scT_{AB}{}^C \nabla_C - \mathscr{R}_{AB}{}^{\underline{c}} X_{\underline{c}} \ .
\end{align}
\end{subequations}
Here the torsion and curvature tensors obey Bianchi identities which follow from
\be 
0 = (-1)^{\ve_A \ve_C} [ \nabla_A , [ \nabla_B , \nabla_C \} \} +
	\text{(two cycles)}\ .
\label{3.15}
\ee
Unlike \eqref{RigidAlgebra}, which is determined by the structure constants,  the graded commutation relations \eqref{gaugedSCA} involve
structure functions $\scT_{AB}{}^C $ and $ \mathscr{R}_{AB}{}^{\underline{c}} $.
Such an algebraic structure is sometimes called a  \emph{soft algebra}, see e.g. \cite{FVP}.


\subsection{Conventional constraints for Weyl multiplet} 
\label{Appendix A.1}

The framework described in the previous subsection 
defines a geometric set-up to obtain a multiplet of conformal supergravity containing the metric. However, in general, the resulting multiplet is reducible. To obtain the irreducible multiplet described in section \ref{section2}, it is necessary to impose constraints on the torsion and curvatures appearing in eq.~\eqref{3.9}. This is a standard task in geometric superspace approaches to supergravity, and it is pedagogically reviewed in \cite{BK,GGRS}. One beautiful feature of the construction of \cite{ButterN=1} is the simplicity of the superspace constraints needed to obtain the Weyl multiplet of conformal supergravity. In fact, to obtain a sufficient set of constraints, one requires the algebra \eqref{3.9} to have a Yang-Mills structure. Specifically, following \cite{ButterN=1} one imposes 
\begin{subequations}
\label{CSSAlgebra-0}
\bea
&	\{ \nabla_{\a} , \nabla_{\b} \}  =  0 ~, \quad \{ \bar{\nabla}_{\ad} , \bar{\nabla}_{\bd} \} = 0 ~, \quad \{\nabla_{\a} , \bar{\nabla}_{\ad} \} = - 2 \ri \nabla_{\a \ad} ~, 
	\\
&\big[ \nabla_{\a} , \nabla_{\b \bd} \big] =  2\ri \ve_{\a \b} \bar{\cal{W}}_\bd
~,\quad
\big[ \bar{\nabla}_{\ad} , \nabla_{\b \bd} \big] =  - 2\ri \ve_{\ad \bd} \cal{W}_{\b}
~,
\eea
\esubeq
where the operator $\bar{\cal{W}}_{\ad}$ is the complex conjugate of $\cal{W}_{\a}$. The latter takes the form
\bea
{\cal{W}}_{\a}
&=&
\frac{1}{2}{\cal{W}}(M)_{\a}{}^{cd} M_{cd}
+{\cal{W}}(\mathbb D)_{\a}\mathbb D
+\ri {\cal{W}}(\mathbb{Y})_{\a} \mathbb{Y}
\non\\
&&
+{\cal{W}}(S)_{\a}{}_\g S^\g
+{\cal{W}}(S)_{\a}{}^\gd \bar{S}_\gd
+{\cal{W}}(K)_{\a}{}_c K^c
~.
\eea
Having imposed the constraints \eqref{CSSAlgebra-0}, the Bianchi identities \eqref{3.15} become non-trivial and now play the role of consistency conditions which may be used to determine the torsion and curvature. Their solution is as follows
\begin{subequations}
	\label{CSSAlgebra}
	\bea
	\big[ \nabla_{\a} , \nabla_{\b \bd} \big] & = & \ri \ve_{\a \b} \Big( 2 \bar{W}_{\bd \gd \dd} \bar{M}^{\gd \dd} - \frac{1}{2} \bar{\nabla}^{\ad} \bar{W}_{\ad \bd \gd} \bar{S}^{\gd} + \frac{1}{2} \nabla_{\g}{}^{\ad} \bar{W}_{\ad \bd \gd} K^{\g \gd} \Big) ~, \\
	\big[ \bar{\nabla}_{\ad} , \nabla_{\b \bd} \big] & = & - \ri \ve_{\ad \bd} \Big( 2 W_{\b}{}^{\g \d} M_{\g \d} + \frac{1}{2} \nabla^{\a} W_{\a \b \g} S^{\g} + \frac{1}{2} \nabla^{\a \gd} W_{\a \b}{}^{\g} K_{\g \gd} \Big) ~,~~~~~~
	\\
	\big[ \nabla_{\a \ad} , \nabla_{\b \bd} \big] & = & \ve_{\ad \bd} \psi_{\a \b} + \ve_{\a \b} \bar{\psi}_{\ad \bd} ~, 
\eea
where the symmetric bispinor operator $\psi_{\a\b}$ and its conjugate $\bar{\psi}_{\ad\bd}$ are given by
	\bea
	\psi_{\a \b} & = & W_{\a \b}{}^{\g} \nabla_{\g} + \nabla^{\g} W_{\a \b}{}^{\d} M_{\g \d} - \frac{1}{8} \nabla^{2} W_{\a \b \g} S^{\g} + \frac{\ri}{2} \nabla^{\g \gd} W_{\a \b \g} \bar{S}_{\gd} \non \\
	&& + \frac{1}{4} \nabla^{\g \dd} \nabla_{(\a} W_{\b) \g}{}^{\d} K_{\d \dd} + \frac{1}{2} \nabla^{\g} W_{\a \b \g} \mathbb{D} - \frac{3}{4} \nabla^{\g} W_{\a \b \g} \mathbb{Y} ~, \\
	\bar{\psi}_{\ad \bd} & = & - \bar{W}_{\ad \bd}{}^{\gd} \bar{\nabla}_{\gd} - \bar{\nabla}^{\gd} \bar{W}_{\ad \bd}{}^{\dd} \bar{M}_{\gd \dd} + \frac{1}{8} \bar{\nabla}^{2} \bar{W}_{\ad \bd \gd} \bar{S}^{\gd} + \frac{\ri}{2} \nabla^{\g \gd} \bar{W}_{\ad \bd \gd} S_{\g} \non \\
	&& - \frac{1}{4} \nabla^{\d \gd} \bar{\nabla}_{(\ad} \bar{W}_{\bd) \gd}{}^{\dd} K_{\d \dd} - \frac{1}{2} \bar{\nabla}^{\gd} \bar{W}_{\ad \bd \gd} \mathbb{D} - \frac{3}{4} \bar{\nabla}^{\gd} \bar{W}_{\ad \bd \gd} \mathbb{Y} ~.
	\eea
\end{subequations}

The structure of the conformal superspace algebra leads to highly non-trivial implications. In particular, eq. \eqref{Knabla.c} implies that 
primary covariantly chiral superfields can carry only undotted spinor indices.
Given such a superfield, $\f_{\a(n)}$, eq. \eqref{Knabla.c} further implies that 
the  $\sU(1)_R$ charge of $\f_{\a(n)}$ 
is determined in terms of its dimension, 
\bea
K^B \f_{\a(n)} =0~, \quad \bar \nabla^\bd \f_{\a(n)} =0 \ ,\quad {\mathbb D} \f_{\a(n)} = w \f_{\a(n)} ~~
\implies ~~ c = -\frac 23  w~.
\label{chirals} 
\eea
There is a regular procedure to construct such constrained multiplets. Given a complex tensor superfield $\j_{\a(n)}$ with the superconformal properties
\bea
K^B \j_{\a(n)} =0~,\quad
{\mathbb D} \j_{\a(n)} = (w-1) \j_{\a(n)}
~,\quad
\mathbb{Y} \psi_{\a(n)} = 2 \Big(1-\frac 13  w\Big) \psi_{\a(n)} 
~,~~~~
\label{chiral-prepotential}
\eea
its descendant 
\bea
\f_{\a(n)}= - \frac 14 \bar \nabla^2 \psi_{\a(n)}
\label{N=2chiralPsi}
\eea
proves to be primary and  covariantly chiral. Here $\f_{\a(n)}$ is invariant under gauge transformations of the form $\d \psi_{\a(n)} = \bar \nabla_\bd \l_{\a(n)}{}^\bd $, where the gauge parameter $\l_{\a(n)}{}^\bd $ is primary.

We note that the conformal superspace algebra is expressed in terms of a single superfield $W_{\alpha \beta \gamma}= W_{(\a\b\g)}$, its conjugate $\bar W_{\ad \bd \gd}$, and their covariant derivatives. This superfield defines an $\mathcal{N}=1$ extension of the Weyl tensor; it is known as the super-Weyl tensor. Further, it is a primary chiral superfield of dimension 3/2 
\bea
\label{SuperWeyl}
K^D W_{\a \b \g} =0~, \quad \bar \nabla^\bd W_{\a\b\g}=0 ~, \quad
{\mathbb D} W_{\a\b\g} = \frac 32 W_{\a\b\g}~,
\eea
and it obeys the Bianchi identity
\bea
B_{\a\ad} :=  \ri \nabla^\b{}_{\ad} \nabla^\g W_{\a\b\g}
=\ri \nabla_{\a}{}^{ \bd} \bar \nabla^\gd \bar W_{\ad\bd\gd}
= \bar B_{\a\ad}~.
\label{super-Bach}
\eea
Here we have defined the superfield $B_{\a\ad}$, which is
the $\cN=1$ supersymmetric generalisation of the Bach tensor introduced in \cite{BK88} (see also \cite{KMT}).
The super-Bach tensor, $B_{\a\ad}$, proves to be primary, $K^B B_{\a\ad} =0$, carries weight $3$, ${\mathbb D} B_{\a\ad} = 3 B_{\a\ad}$, and satisfies the conservation equation
\bea
\nabla^\a B_{\a\ad}=0 ~~~\Longleftrightarrow~~~
\bar \nabla^\ad B_{\a\ad} =0
~.
\eea

At this stage, it is still necessary to show that the conformal superspace geometry described in this subsection indeed 
encodes the Weyl multiplet of conformal supergravity. We will demonstrate this in two different ways. First, we will describe the procedure of reducing the results of this section to their component field description. Secondly, we will prove that this geometry is equivalent to the $\sU(1)$, and consequently the Grimm-Wess-Zumino, superspace descriptions of conformal supergravity \cite{Howe,GWZ}.


\subsection{Superconformal action principles}

In order to formulate locally superconformal field theories, an action principle is required. 
As in the rigid supersymmetric case, locally superconformal actions can be constructed in two different ways:  either as integrals over the full superspace or over its chiral subspace. Here we consider separately these two options. 
  
We look for a scalar superfield $\cL$ such that the action
\be
\label{3.24}
S = \int\text{d}^{4|4}z \, E \, \mathcal{L} ~, \qquad \rd^{4|4}z := \rd^{4}x \, \rd^{2} \q \, \rd^{2} \bar{\q}~
\ee
is locally superconformal.  
Performing a  gauge transformation, eqs. \eqref{SUGRAtransmations} and \eqref{3.14}, we arrive at the variation
\bea
\label{3.25}
\d_{\mathcal{K}} S = \int\text{d}^{4|4}z \, E \, \bigg ( (-1)^{\ve_A}\Big[  \nabla_A( \xi^A \mathcal{L}) &+&  
\xi^B   \scT_{BA}{}^A  \cL \Big]\non \\ 
 \qquad + \L^{\underline{b}}\Big[  (-1)^{\ve_A}  f_{\underline{b} A}{}^{A} \mathcal{L} 
 &+& 
  X_{\underline{b}} \mathcal{L}\Big] \bigg) ~, 
\eea
which must vanish.  
Now, requiring the contributions containing $\L^{\underline{b}}$ to vanish gives
\be
\label{3.27}
X_{\underline{b}} \mathcal{L} = - (-1)^{\ve_A} f_{\underline{b} A}{}^{A} \mathcal{L}
\quad \Longleftrightarrow \quad K^B \cL =0~, \quad {\mathbb D} \cL = 2 \cL~, 
\quad \mathbb{Y} \cL =0~.
\ee
Once the conditions 
\eqref{3.27} are satisfied, it is readily seen that the remaining $\x$-dependent contributions in \eqref{3.25} cancel out. In summary, given a primary real dimension-2 scalar Lagrangian $\cL$, the action \eqref{3.24} is locally superconformal. 

Given a  primary chiral scalar Lagrangian $\cL_{\rm c}$ of weight $+3$, 
\bea
K^B \cL_{\rm c} =0~, \quad \bar \nabla_\ad \cL_{\rm c} =0 \ ,\quad 
{\mathbb D} \cL_{\rm c} = 3 \cL_{\rm c} ~,
\label{ChiralLagrangian}
\eea
the  {\rm chiral} action
\bea
S_{\rm c}=\int\rd^4x\rd^2\q\, \cE \,\cL_{\rm c} 
\label{chiralAc}
\eea
is locally superconformal. Here $\cE$ is a chiral density. The precise definition of $\cE$ requires the use of a prepotential formulation for supergravity, see subsection \ref{section5.5}.

A different definition of $S_{\rm c}$ exists, which is based on the use of a complex superfield $\U$ with the following superconformal properties (for some constant $\D$)
\bea 
K^B \U =0~, \quad 
{\mathbb D} \U = (\D-1) \U ~,\quad \mathbb{Y} \U = 2\big(1-\frac 13 \D\big) \U~, 
\eea
such that $\bar \nabla^2 \U$ is nowhere vanishing,  that is $(\bar \nabla^2 \U )^{-1}$ exists.
Specifically, the chiral action may be identified with the functional 
\bea
S_{\rm c} = -4  \int\text{d}^{4|4}z \, E \, \frac{\U}{\bar \nabla^2 \U}
\mathcal{L}_{\rm c} ~,
\label{3.30}
\eea
which 
possesses the two fundamental properties:
(i) it is locally superconformal under the conditions \eqref{ChiralLagrangian};
and (ii) it is independent of $\U$, 
\bea
\d_\U  \int\text{d}^{4|4}z \, E \, \frac{\U}{\bar \nabla^2 \U}
\mathcal{L}_{\rm c} =0~,
\eea
for an arbitrary variation $\d \U$. Using the representation \eqref{3.30} for the chiral action \eqref{chiralAc}, it holds that
\bea
\int\text{d}^{4|4}z \, E \, \mathcal{L}  = \int\rd^4x\rd^2\q\, \cE \,\cL_{\rm c} ~, \qquad 
\cL_{\rm c} = - \frac 14 \bar \nabla^2 \cL~.
\eea
This result can also be obtained using superspace normal coordinates \cite{ButterN=1}, see also 
\cite{KT-M2009}.

There is an alternative definition of the chiral action that follows from the superform approach to 
the construction of supersymmetric invariants \cite{Ectoplasm,GGKS}.
It is based on the use of the  following  super 4-form 
\bea
\Xi_4&=&
2\ri \bar{E}_\dd \wedge \bar{E}_\gd\wedge  E^b\wedge  E^a(\ts_{ab})^{\gd\dd}
\cL_{\rm c}
+\frac{\ri}{6}\ve_{abcd}\bar{E}_\dd\wedge  E^c\wedge E^b\wedge  E^a (\ts^d)^{\dd\d}\nabla_\a\cL_{\rm c}
\non\\
&&
-\frac{1}{96}\ve_{abcd}E^d\wedge E^c\wedge E^b\wedge E^a 
\nabla^2\cL_{\rm c}~,
\label{Sigma_4}
\eea
which was constructed by Bin\'etruy {\it et al.} 
\cite{Binetruy:1996xw} and independently by Gates {\it et al.} \cite{GGKS} in the GWZ superspace. This superform is closed, 
 \bea
 \rd \, \X_4 =0~.
 \eea
 It proves to be primary\footnote{The superform may be degauged to the GWZ superspace described in the next section. Then the condition \eqref{337} is equivalent to the super-Weyl invariance of $\X_4$. The latter property was proved in \cite{KT-M17}. } 
 \bea
 K^B \X_4 =0~.
 \label{337}
 \eea
 The chiral action \eqref{chiralAc} can be recast
  as an integral of $\Xi_4$ over a spacetime $\cM^4$,
\begin{subequations}\label{2.24}
 \bea
 S_{\rm c} &=& \int_{\cM^4} \Xi_4
 ~, \label{2.24.a}
\eea
where $\cM^4$ is the bosonic body of the curved superspace  $\cM^{4|4}$
obtained by switching off  the Grassmann variables. 
 It turns out that  \eqref{2.24.a} leads to the representation 
  \bea
 S_{\rm c} = \int {\rm d}^4x\, e \,\Big(
-\frac{1}{4}\de^2
+\frac{\ri}{2}(\tilde{\s}^a)^{\ad\a}\bar{\Psi}_a{}_\ad\de_\a
-(\tilde{\s}^{ab})^{\ad\bd}\bar{\Psi}_a{}_\ad\bar{\Psi}_b{}_\bd\Big){\cL}_{\rm c}\Big|_{\theta=0} ~~~~~~~
\label{2.24.b}
\eea
\end{subequations}
which is the simplest way to reduce the action from superfields to components. 
Here $\bar{\Psi}_{a \ad } = e_a{}^m \bar \J_{m\ad} $ is the gravitino defined in
\eqref{WeylMult}.


\subsection{Component reduction and the Weyl multiplet} \label{section3.4}

Having  formulated conformal superspace in the preceding subsections, it is instructive
to utilise it in reproducing the  results of section \ref{section2.3}. Specifically, we will
briefly describe the Weyl multiplet of conformal supergravity 
(see e.g. \cite{FVP,FT} for pedagogical reviews) 
and $Q$-supersymmetry transformations of the corresponding fields. As described earlier, the former involves a set of gauge one-forms: the vielbein $e_m{}^a$, gravitino $\Psi_m{}^\a$, $\sU(1)_R$ gauge field $\mathfrak{A}_m$ and dilatation gauge field $b_m$. 
Modulo purely gauge degrees of freedom, 
they may be shown to be the only independent geometric fields and arise as the lowest components of the superforms
\begin{align}
	\label{WeylMult}
	e_m{}^a := E_m{}^a|~, \qquad \Psi_m{}^\a := 2 E_m{}^\a| ~, \qquad \mathfrak{A}_m := \Phi_m | ~, \qquad b_m := B_m |~,
\end{align}
where the bar projection of a superfield $\Xi(x,\q,\bar{\q})$ is defined by $\Xi| := \Xi|_{\q = \bar{\q} = 0}$. 

It remains to compute the $Q$-supersymmetry transformations of the fields \eqref{WeylMult} and show that they do indeed form the Weyl multiplet. By employing \eqref{3.8}, their transformations when $\mathcal{K}(\xi) = \xi^\a \nabla_\a + \bar{\xi}_\ad \bar{\nabla}^{\ad}$ may be shown to be:
\begin{subequations}
	\label{3.40}
	\begin{align}
		\d_{\mathcal{K}(\ve)} e_m{}^a &= \ri ( \ve \s^a \bar{\Psi}_m - \Psi_m \s^a \bar{\ve} ) ~, \\
		\d_{\mathcal{K}(\ve)} \Psi_m{}^\a &= 2 \hat{\nabla}_m \ve^\a ~, \\
		\d_{\mathcal{K}(\ve)} \mathfrak{A}_m &= 3 \ri ( \mathfrak{F}_m{}^\a | \ve_\a - \bar{\mathfrak{F}}_{m \ad} | \bar{\ve}^\ad) ~, \\
		\d_{\mathcal{K}(\ve)} b_m &= 2 ( \mathfrak{F}_m{}^\a | \ve_\a + \bar{\mathfrak{F}}_{m \ad} | \bar{\ve}^\ad) ~,
	\end{align}
\end{subequations}
where we have denoted $\ve^\a := \xi^\a|$ and $\hat{\nabla}_m$ was defined in \eqref{2.44}. Additionally, by a routine analysis of the spinor torsion two-form $\mathcal{T}_{mn}{}^\a|$, it may be shown that 
\begin{subequations}
\begin{align}
	(\s^{mn})^{\b \g} \hat{\nabla}_{m} \Psi_n{}^\a + 2 \ri \ve^{\a (\b} (\s^m)^{\g)}{}_{\ad} \bar{\mathfrak{F}}_m{}^\ad| = W^{\a \b \g}|~,
\end{align}
which implies:
\begin{align}
	W_{\a \b \g} | &= (\s^{mn})_{(\a \b} \hat{\nabla}_{m} \Psi_{n \g)}~, \\
	\label{3.41c}
	\mathfrak{F}_m{}^\a | &= \frac{\ri}{3} \hat{\nabla}_{[m} \bar{\Psi}_{n] \ad} (\tilde{\s}^n)^{\ad \a} - \frac{1}{12} g_{mn} \ve^{n i j k} \hat{\nabla}_i \bar{\Psi}_{j \ad} (\tilde{\s}_k)^{\ad \a} ~.
\end{align}
\end{subequations}
Then, upon inserting \eqref{3.41c} into \eqref{3.40}, the transformations coincide with \eqref{2.43}. 

To conclude, it should be noted that the dilatation gauge field $b_m$ describes purely gauge degrees of freedom. This may be seen by noting that, according to \eqref{3.8}, it transforms algebraically when $\mathcal{K}(\L) = \L_a K^a$
\begin{align}
	\label{3.39}
	\d_{\mathcal{K}(\L)} b_m = - 2 \L_m | ~.
\end{align}
Hence, we impose the gauge $b_m = 0$ by fixing the special conformal gauge freedom. It should also be noted that the remaining fields appearing in \eqref{WeylMult} are inert under such transformations. Additionally, in order to preserve the gauge $b_m = 0$,  each
$Q$-supersymmetry transformation \eqref{3.40} must be accompanied 
with a compensating special conformal transformation \eqref{3.39} with $\L_m (\ve) = \mathfrak{F}_m{}^\a | \ve_\a + \bar{\mathfrak{F}}_{m \ad} | \bar{\ve}^\ad$. 
As a result, we have shown that the fields $\big\{ e_m{}^a , \J_{m\a} , \bar \J_{m}^\ad , \mathfrak{A}_m \big\}$ do indeed constitute the reduced Weyl multiplet introduced in subsection 
\ref{section2.3}.


\section{Other superspace formulations for conformal supergravity}\label{Section4}

As pointed out in section \ref{Section1}, conformal superspace is not the only superspace setting to describe conformal supergravity. The other most popular formulations are: (i) $\sU(1)$ superspace \cite{Howe}; and (ii) the GWZ superspace \cite{GWZ}. They differ by their structure groups, which are $\sSL( 2, {\mathbb C}) \times \sU(1)_R$ and $\sSL( 2, {\mathbb C}) $, respectively. Both of them can be derived from conformal superspace. Below we describe 
the relevant degauging procedures.

\subsection{The $\sU(1)$ superspace geometry}
\label{DegaugingSection}

According to \eqref{SUGRAtransmations}, under an infinitesimal special superconformal gauge transformation $\mathcal{K} = \Lambda_{B} K^{B}$, the dilatation connection transforms as follows
\bea
\d_{\mathcal{K}} B_{A} = - 2 \Lambda_{A} ~.
\eea
Thus, it is possible to choose a gauge condition $B_{A} = 0$, which completely fixes 
the special superconformal gauge freedom.\footnote{There is a class of residual gauge transformations preserving the gauge $B_{A}=0$. These generate the super-Weyl transformations of $\sU(1)$ superspace, see the next subsection.} As a result, the corresponding connection is no longer required for the covariance of $\nabla_A$ under the residual gauge freedom and
may be extracted from $\nabla_{A}$,
\bea
\nabla_{A} &=& \mathfrak{D}_{A} - \mathfrak{F}_{AB} K^{B} ~. \label{ND}
\eea
Here the operator $\mathfrak{D}_{A} $ involves only the Lorentz and $\sU(1)_R$ connections.

The next step is to relate the special superconformal connection
$\mathfrak{F}_{AB}$  to the torsion tensor of $\sU(1)$ superspace. To do this, it is necessary to make use of the relation
\bea
\label{4.3}
[ \nabla_{A} , \nabla_{B} \} &=& [ \mathfrak{D}_{A} , \mathfrak{D}_{B} \} - \big(\mathfrak{D}_{A} \mathfrak{F}_{BC} - (-1)^{AB} \mathfrak{D}_{B} \mathfrak{F}_{AC} \big) K^C - \mathfrak{F}_{AC} [ K^{C} , \nabla_B \} \non \\
&& + (-1)^{AB} \mathfrak{F}_{BC} [ K^{C} , \nabla_A \} + (-1)^{BC} \mathfrak{F}_{AC} \mathfrak{F}_{BD} [K^D , K^C \} ~.
\eea
In conjunction with \eqref{CSSAlgebra}, this relation leads to a set of consistency conditions that are equivalent to the Bianchi identities of $\sU(1)$ superspace \cite{Howe}. 
Their solution expresses the components of $\mathfrak{F}_{AB}$ in terms of the torsion 
tensor of $\sU(1)$ superspace. 
We will not provide a detailed analysis for this step and instead refer the reader to the proof in \cite{ButterN=1}. The outcome of the analysis is as follows:
\begin{subequations} \label{connections}
	\bea
	\mathfrak{F}_{\a \b} & = & \frac{1}{2} \ve_{\a \b} \bar{R} ~, \quad \bar{\mathfrak{F}}_{\ad \bd} = -\frac{1}{2} \ve_{\ad \bd} R ~, \quad
	\mathfrak{F}_{\a \bd} 
	= -\bar{\mathfrak{F}}_{\bd \a} 
	= \frac{1}{4} G_{\a \bd} ~, 
		 \\
	\mathfrak{F}_{\a , \b \bd} & = & - \frac{\ri}{4} \mathfrak{D}_{\a} G_{\b \bd} - \frac{\ri}{6} \ve_{\a \b} \bar{X}_{\bd}  = \mathfrak{F}_{\b \bd , \a} ~, \\
	\bar{\mathfrak{F}}_{\ad , \b \bd} &=& \frac{\ri}{4} \bar{\mathfrak{D}}_{\ad} G_{\b \bd} + \frac{\ri}{6} \ve_{\ad \bd} X_{\b} =\mathfrak{F}_{\b \bd , \a} ~, \\
	\mathfrak{F}_{\a \ad , \b \bd} & = & - \frac{1}{8} \big[ \mathfrak{D}_{\a} , \bar{\mathfrak{D}}_{\ad} \big] G_{\b \bd} - \frac{1}{12} \ve_{\ad \bd} \mathfrak{D}_{\a} X_{\b} + \frac{1}{12} \ve_{\a \b} \bar{\mathfrak{D}}_{\ad} \bar{X}_{\bd}  \non \\
	&& 
	+ \frac{1}{2} \ve_{\a \b} \ve_{\ad \bd} \bar{R} R
	+ \frac{1}{8} G_{\a \bd} G_{\b \ad} ~,
	\eea
\end{subequations}
where $R$ and $X_{\a}$ are complex chiral
\begin{subequations}
\bea
\bar{\mathfrak{D}}_{\ad} R &=& 0 ~, \qquad \mathbb{Y} R = -2 R~, 
\\
\bar{\mathfrak{D}}_{\ad} X_{\a} &=& 0 ~, \qquad \mathbb{Y} X_\a = - X_\a~,
\eea
while $G_{\a \ad}$ is a real vector superfield. These are related via
\bea
X_{\a} &=& \mathfrak{D}_{\a}R - \bar{\mathfrak{D}}^{\ad}G_{\a \ad} ~. \label{Bianchi1}
\eea
\end{subequations}

We now pause and comment on the geometry described by $\mathfrak{D}_A$. In particular, by employing \eqref{4.3} one arrives at the following anti-commutation relation
\be
\label{4.6}
\{ \mathfrak{D}_\a , \bar{\mathfrak{D}}_{\ad} \} = - 2 \ri \mathfrak{D}_{\aa} - G^{\b}{}_{\ad} M_{\a \b}
 + G_\a{}^{\bd} \bar{M}_{\ad \bd} + \frac{3}{2} G_\aa \mathbb{Y} ~.
\ee
With the goal of simplicity in performing calculations within $\sU(1)$ superspace, we prefer to work with a geometry where the right hand side of \eqref{4.6} contains no curvature-dependent terms. To this end, we perform the following redefinition
\begin{subequations}
\label{A.16}
\bea
\mathfrak{D}_{\a \ad} = \cD_{\a \ad} + \frac{\ri}{2} G^{\b}{}_{\ad} M_{\a \b} &-& \frac{\ri}{2} G_{\a}{
}^{\bd} \bar{M}_{\ad \bd} - \frac{3 \ri}{4} G_{\a \ad} \mathbb{Y} ~,  \\
\mathfrak{D}_{\a} = \cD_{\a} ~, \, && \, \bar{\mathfrak{D}}^{\ad} = \bar{\cD}^{\ad} ~,
\eea
\end{subequations}
where $\cD_{A}$ takes the form\bea
\cD_{A} & = & E_{A} - \frac{1}{2}  \hat{\O}_{A}{}^{bc} M_{bc} - {\rm i}\,  \hat{\F}_{A} {\mathbb Y} \non \\
& = & E_{A} -  \hat{\O}_{A}{}^{\b \g} M_{\b \g} -  \hat{\bar{\O}}_{A}{}^{\bd \gd} \bar{M}_{\bd \gd} - {\rm i}\,  \hat{\F}_{A} {\mathbb Y} ~.
\eea
Here we have attached a hat to each connection superfield to distinguish them from their cousins residing in the conformal covariant derivative $\nabla_A$. In what follows, 
these hats will be omitted.  

Now it may be shown that the algebra obeyed by $\cD_A$ takes the form
\begin{subequations} \label{U(1)algebra}
	\bea
	\{ \cD_{\a}, \cD_{\b} \} &=& -4{\bar R} M_{\a \b}~, \qquad
	\{\cDB_{\ad}, \cDB_{\bd} \} =  4R {\bar M}_{\ad \bd}~, \label{U(1)algebra.a}\\
	&& {} \qquad \{ \cD_{\a} , \cDB_{\ad} \} = -2{\rm i} \cD_{\a \ad} ~, 
\label{U(1)algebra.b}	\\
	\big[ \cD_{\a} , \cD_{ \b \bd } \big]
	& = &
	{\rm i}
	{\ve}_{\a \b}
	\Big({\bar R}\,\cDB_\bd + G^\g{}_\bd \cD_\g
	- (\cD^\g G^\d{}_\bd)  M_{\g \d}
	+2{\bar W}_\bd{}^{\gd \dot{\d}}
	{\bar M}_{\gd \dot{\d} }  \Big) \non \\
	&&
	+ {\rm i} (\cDB_{\bd} {\bar R})  M_{\a \b}
	-\frac{\ri}{3} \ve_{\a\b} \bar X^\gd \bar M_{\gd \bd} - \frac{\ri}{2} \ve_{\a\b} \bar X_\bd \mathbb{Y}
	~, \label{U(1)algebra.c}\\
	\big[ {\bar \cD}_{\ad} , \cD_{\b\bd} \big]
	& = &
	- {\rm i}
	\ve_{\ad\bd}
	\Big({R}\,\cD_{\b} + G_\b{}^\gd \cDB_\gd
	- (\cDB^{\gd} G_{\b}{}^{\dd})  \bar M_{\gd \dd}
	+2{W}_\b{}^{\g \d}
	{M}_{\g \d }  \Big) \non \\
	&&
	- {\rm i} (\cD_\b R)  {\bar M}_{\ad \bd}
	+\frac{\ri}{3} \ve_{\ad \bd} X^{\g} M_{\g \b} - \frac{\ri}{2} \ve_{\ad\bd} X_\b \mathbb{Y}
	~, \label{U(1)algebra.d}
	\eea
	which lead to 
	\bea
	\left[ \cD_{\a \ad} , \cD_{\b \bd} \right] & = & \ve_{\a \b} \bar \chi_{\ad \bd} + \ve_{\ad \bd} \chi_{\a \b} ~, \\
	\chi_{\a \b} & = & - \ri G_{ ( \a }{}^{\gd} \cD_{\b ) \gd} + \frac{1}{2} \cD_{( \a } R \cD_{\b)} + \frac{1}{2} \cD_{ ( \a } G_{\b )}{}^{\gd} \cDB_{\gd} + W_{\a \b}{}^{\g} \cD_{\g} \non \\
	&& + \frac{1}{6} X_{( \a} \cD_{ \b)} + \frac{1}{4} (\cD^{2} - 8R) {\bar R} M_{\a \b} + \cD_{( \a} W_{ \b)}{}^{\g \d} M_{\g \d} \non \\
	&& - \frac{1}{6} \cD_{( \a} X^{\g} M_{\b) \g} - \frac{1}{2} \cD_{ ( \a} \cDB^{\gd} G_{\b)}{}^{\dd} {\bar M}_{\gd \dd} + \frac{1}{4} \cD_{( \a} X_{\b)} \mathbb{Y} ~, \\
	{\bar \chi}_{\ad \bd} & = & \ri G^{\g}{}_{( \ad} \cD_{\g \bd)} - \frac{1}{2} \cDB_{( \ad } {\bar R} \cDB_{\bd)} - \frac{1}{2} \cDB_{ ( \ad } G^{\g}{}_{\bd)} \cD_{\g} - {\bar W}_{\ad \bd}{}^{\gd} \cDB_{\gd} \non \\
	&& - \frac{1}{6} {\bar X}_{( \ad} \cDB_{ \bd)} + \frac{1}{4} (\cDB^{2} - 8{\bar R}) R {\bar M}_{\ad \bd} - \cDB_{( \ad} {\bar W}_{ \bd)}{}^{\gd \dd} {\bar M}_{\gd \dd} \non \\
	&& + \frac{1}{6} \cDB_{( \ad} {\bar X}^{\gd} {\bar M}_{\bd) \gd} + \frac{1}{2} \cDB_{ ( \ad} \cD^{\g} G^{\d}{}_{\bd)} M_{\g \d} + \frac{1}{4} \cDB_{( \ad} {\bar X}_{\bd)} \mathbb{Y} ~.
	\eea
\end{subequations}
These relations should be supplemented with the following Bianchi identities:
\begin{subequations}
\bea
\cD^{\a} X_{\a} &=& \cDB_{\ad} {\bar X}^{\ad} ~, \label{Bianchi2}\\
\bar{\cD}_{\ad} W_{\a \b \g} &=& 0 ~, \\
\cD^{\g} W_{\a \b \g} &=& {\rm i} \cD_{(\a}{}^{\gd} G_{\b ) \gd} - \frac{1}{3} \cD_{(\a} X_{\b)} ~.
\label{4.10c}
\eea
\end{subequations}
In particular, it should be noted that \eqref{Bianchi2} implies $X_{\a}$ is the chiral field strength of a $\sU(1)$ vector multiplet. The geometry described above is the $\sU(1)$ superspace geometry \cite{Howe,GGRS} in the form described in \cite{KR,BK11}.

To conclude our discussion of the $\sU(1)$ superspace geometry, we make two comments. 
Firstly, one may check that degauging the relation \eqref{N=2chiralPsi} gives
\bea
\f_{\a(n)}= - \frac 14 \big(\bar \cD^2 -4R \big)\psi_{\a(n)}~, \qquad 
\bar \cD_\bd \f_{\a(n)}=0~.
\eea
Secondly, integration by parts is remarkably simple in $\sU(1)$ superspace:
\bea
\int\text{d}^{4|4}z \, E \,  (-1)^{\ve_A} \cD_A \cV^A =0~,
\eea
where $\cV^A $ is arbitrary.\footnote{In conformal superspace, 
integration by parts requires special care \cite{ButterN=1}.}


\subsection{The super-Weyl transformations of $\sU(1)$ superspace}

In the previous subsection we made use of the special conformal gauge freedom to degauge from conformal to $\sU(1)$ superspace. Here we will show that the residual dilatation symmetry manifests in the latter as super-Weyl transformations.

Specifically, to preserve the gauge $B_{A}=0$, every local dilatation transformation with parameter $ \S $ should be accompanied by a compensating special conformal one, $\L^{B} (\S)$
\begin{align}
\mathcal{K}(\Sigma) = \L_{B} (\Sigma)K^{B} + \S \mathbb{D} \quad \implies \quad \d_{\mathcal{K}( \S )} B_A = 0~.
\end{align}
We then arrive at the following constraints
\bea
\L_{A}(\S) = \hf \nabla_A \S~.
\eea
As a result, we define the following transformation
\bea
\d_{\S} \nabla_{A} &=& \d_{\S} \mathfrak{D}_{A} - \d_{\S} \big(\mathfrak{F}_{AB} K^{B} \big)
= [\mathcal{K}(\S) ~, \nabla_{A} ]~.
\eea

By making use of \eqref{A.16} and \eqref{connections}, we arrive at the following transformation laws for the $\sU(1)$ superspace covariant derivatives
\begin{subequations}
\label{superWeylTf}
\bea
\delta_{\S}\cD_{\a} & = & \frac{1}{2} \S \cD_{\a} + 2 \cD^{\b} \Sigma M_{\b \a} - \frac{3}{2} \cD_{\a} 
\Sigma \mathbb{Y} ~, \\
\d_{\S} \cDB_{\ad} & = & \frac{1}{2} \S \cDB_{\ad} + 2 \cDB^{\bd} \S {\bar M}_{\bd \ad} +
\frac{3}{2} \cDB_{\ad} \S \mathbb{Y} ~, \\
\d_{\S} \cD_{\a \ad} & = & \S \cD_{\a \ad} + {\rm i} \cD_{\a} \S \cDB_{\ad} 
+ {\rm i} \cDB_{\ad} \S \cD_{\a}  + {\rm i} \cDB_{\ad} \cD^{\b} \S  M_{\b \a} \non \\
&& + {\rm i} \cD_{\a} \cDB^{\bd} \S { \bar M}_{\bd \ad} + \frac{3}{4} {\rm i}  \left[ \cD_{\a} , \cDB_{\ad} \right]\S \mathbb{Y} ~, \label{superWeylTf.c}
\eea
\end{subequations}
while the torsion superfields arising from the degauged torsion $\mathfrak{F}_{AB}$ transform as follows
\begin{subequations}
\label{superWeylTfTorsions}
\bea
\d_{\S} R & = & \S R + \frac{1}{2} \cDB^{2} \S ~, \\
\d_{\S} G_{\a \ad} & = &  \S G_{\a \ad} + [ \cD_{\a} , \cDB_{\ad} ] \S ~, \\
\d_{\S} X_{\a} & = & \frac{3}{2} \S X_{\a} - \frac{3}{2} (\cDB^{2} - 4 R) \cD_{\a} \S ~. \label{sWX}
\eea
Finally, as the super-Weyl tensor is conformally covariant, its transformation law is readily obtained via
\bea
\d_{\S} W_{\a \b \g} = \big(\L_{B}(\S) K^{B} + \S \mathbb{D}\big) W_{\a \b \g} = \frac{3}{2} \S W_{\a \b \g}~.
\eea
\end{subequations}

The relations  \eqref{superWeylTf} and \eqref{superWeylTfTorsions} 
give the super-Weyl transformations 
in $\sU(1)$ superspace
\cite{GGRS,Howe} (see also \cite{KR,BK11}).
The conditions \eqref{PrimaryDimensionCharge}, which define a primary superfield $U$, are equivalent to the following 
\bea
\d_\S U = w \S U~, \qquad  {\mathbb Y} U = c U~,
\label{primaryU(1)}
\eea
in $\sU(1)$ superspace. 

The $\sU(1)$ superspace formulation was fully developed in the book \cite{GGRS} in which various applications were also given.


\subsection{The Grimm-Wess-Zumino formulation}

As pointed out in section \ref{DegaugingSection}, the covariantly chiral spinor $X_\a$ is the field strength 
of an Abelian vector multiplet. It follows from \eqref{sWX}
that the super-Weyl gauge freedom allows us to choose the gauge 
\bea
X_\a =0~.
\eea
In this gauge the $\sU(1)_{R}$ curvature vanishes, in accordance with  
\eqref{U(1)algebra}, and therefore the $\sU(1)_{R}$ connection may be gauged away, 
\bea
{\F}_A =0~.
\eea
Then, the algebra of covariant derivatives \eqref{U(1)algebra} 
reduces to that describing the  GWZ geometry \cite{GWZ}.

Equation \eqref{sWX}  tells us that imposing the condition $X_\a=0$ does not fix completely the super-Weyl freedom. The residual transformations are generated 
by parameters of the form 
\bea 
\S=\hf \big(\s +\bar \s \big) ~, \qquad \bar \cD_\ad \s =0~.
\label{2.18}
\eea
However, in order to preserve the  $\sU(1)_{R}$ gauge ${\F}_A=0$, 
every residual super-Weyl transformation \eqref{2.18} must be accompanied by a 
compensating $\sU(1)_{R}$ transformation with 
\bea
\r = \frac{3}{4} \ri \big( \s - \bar \s\big)~.
\eea
This leads to the transformation \cite{HT} 
\begin{subequations} 
\label{superweyl}
\bea
\d_\s \cD_\a &=& ( {\bar \s} - \hf \s)  \cD_\a + (\cD^\b \s) \, M_{\a \b}  ~, \\
\d_\s \bar \cD_\ad & = & (  \s -  \hf {\bar \s})
\bar \cD_\ad +  ( \bar \cD^\bd  {\bar \s} )  {\bar M}_{\ad \bd} ~,\\
\d_\s \cD_{\a\ad} &=& \hf( \s +\bar \s) \cD_{\a\ad} 
+\frac{\ri}{2} (\bar \cD_\ad \bar \s) \cD_\a + \frac{\ri}{2} ( \cD_\a  \s) \bar \cD_\ad \non \\
&& + (\cD^\b{}_\ad \s) M_{\a\b} + (\cD_\a{}^\bd \bar \s) \bar M_{\ad \bd}~.
\eea
\end{subequations}
The torsion tensors  transform
 as follows:
\begin{subequations} 
\bea
\d_\s R &=& 2\s R +\frac{1}{4} (\bar \cD^2 -4R ) \bar \s ~, \\
\d_\s G_{\a\ad} &=& \hf (\s +\bar \s) G_{\a\ad} +\ri \cD_{\a\ad} ( \s- \bar \s) ~, 
\label{s-WeylG}\\
\d_\s W_{\a\b\g} &=&\frac{3}{2} \s W_{\a\b\g}~.
\label{s-WeylW}
\eea
\end{subequations} 
The conditions \eqref{primaryU(1)}, which define a primary superfield $U$, turn into 
\bea
\d_\s U = (p\s +q\bar \s) U~, \qquad p+q =w, \quad p-q = - \frac 32 c~,
\label{primaryGWZ}
\eea
in the GWZ approach.

Let us fix a background curved superspace   $(\cM^{4|4}, \cD)$. 
A supervector field $\x= \x^B E_B$ on this superspace
is called {\it conformal Killing} if there exists a 
Lorentz parameter $K^{bc}[\x] $ and a super-Weyl chiral  parameter $\s [\x]$ such that 
\bea
\big[ \x^B\cD_B + \hf K^{bc} [\x] M_{bc} , \cD_A\big]  + \d_{\s[\x]} \cD_A =0~.
\label{conf_Killing}
\eea
In other words, the  coordinate transformation generated by $\x$ is accompanied by certain Lorentz
and super-Weyl transformations such that the superspace geometry does not change. It can be shown \cite{BK} that the equation \eqref{conf_Killing} uniquely determines the spinor components of 
$\x^B= (\x^b, \x^\b , \bar \x_\bd) $ and the parameters $K^{bc}[\x] $ and  $\s [\x]$ in terms of $\x^b$, 
and the latter obeys the equation 
\bea
\cD_{(\a} \x_{\b) \bd} =0 \quad \Longleftrightarrow \quad 
\bar \cD_{(\ad} \x_{\b \bd)} =0 ~.
\eea
The set of all conformal Killing supervector fields on $(\cM^{4|4}, \cD)$ constitutes the superconformal algebra of $(\cM^{4|4}, \cD)$. Given a super-Weyl invariant theory on $(\cM^{4|4}, \cD)$ described by primary superfields $U$, its action is invariant under the superconformal transformations
\bea
\d_\x U =  \cK[\x] U~,
\quad \cK[\x] = \x^B\cD_B + \hf K^{bc}[\x] M_{bc}  +  p \s[\x] + q \bar \s[\x] ~,
\label{primaryTL-curved}
\eea
for an arbitrary conformal Killing supervector field $\x$. In the case that  $(\cM^{4|4}, \cD)$
coincides with Minkowski superspace,  $({\mathbb M}^{4|4}, D)$, the superconformal Killing equation 
\eqref{conf_Killing} is equivalent to \eqref{4Dmaster2} and the transformation law  
\eqref{primaryTL-curved}  to \eqref{primaryTL}.

The GWZ formulation has been used in most applications of $\cN=1$ superfield 
supergravity. It is reviewed in several textbook, see, e.g.,  \cite{WB,BK,RauschdeTraubenberg:2020kol}.


\section{Supergravity prepotentials} 

The constraints on the GWZ geometry \cite{WZ,GWZ} were solved by Siegel \cite{Siegel78} in terms of unconstrained prepotentials. This solution was extended to non-minimal  supergravity ($n\neq -1/3,0$) by Gates and Siegel  \cite{SG}, and then 
to $\sU(1)$ superspace in the book \cite{GGRS}. Here we review the original solution given in \cite{Siegel78}. The prepotential description of conformal superspace was worked out in \cite{ButterN=1}, and the interested reader is referred to the original publication.

The covariant derivatives have the form 
\bea
\cD_A= E_A -  \hf \O_A{}^{bc} M_{bc}~
\eea
and obey the graded commutation relations 
\bea
\big[ \cD_A , \cD_B \big\} = - \cT_{AB}{}^C \cD_C - \hf \cR_{AB}{}^{cd} M_{cd} ~,
\eea
where the torsion and curvature tensors are read off from \eqref{U(1)algebra} by setting $X_\a =0$.

The gauge group of conformal supergravity is generated by the general coordinate ($K^N$), local Lorentz ($K^{bc}$) and super-Weyl ($\s$ and $\bar \s$) transformations. The combined general coordinate and Local Lorentz transformation acts on $\cD_A$ and a tensor superfield $\cT$ (with suppressed indices) by the rule 
\bea
\cD'_A &=& \re^\cK \cD_A \re^{-\cK} ~, \qquad \cT' = \re^\cK \cT~, \qquad 
\cK = K^N \pa_N + \hf K^{bc} M_{bc}~.
\label{K-group}
\eea
The super-Weyl transformation of $\cD_A$ is given by eq. \eqref{superweyl}, with the parameter $\s$ being covariantly chiral. 
Given a primary tensor superfield $U$, its super-Weyl transformation law is 
given by eq. \eqref{primaryGWZ}.


\subsection{Spinor covariant derivatives}

Nontrivial information is contained in the relations \eqref{U(1)algebra.a} and
\eqref{U(1)algebra.b}. First of all, the spinor components of the connection
$\O_A{}^{bc} = (\O_a{}^{b c} , \O_\a{}^{bc}, \bar \O^{\ad bc} )$ are determined in terms of  the anholonomy coefficients, $C_{AB}{}^C$, defined by 
\bea
\big[ E_A , E_B\big\} = C_{AB}{}^C E_C~.
\eea 
In particular, for $\hf \O_\a{}^{bc} M_{bc} =\O_\a{}^{\b\g} M_{\b\g} + \O_\a{}^{\bd \gd} \bar M_{\bd \gd} $ we  obtain
\bea
\O_{\a\b\g} = \hf\Big( C_{\a \b\g} +C_{\a \g\b} - C_{\b \g \a}\Big) ~, \qquad 
\O_{\a\bd \gd} = -C_{\a\bd \gd}~.
\eea
Secondly, since the curvature $R_{\a\b}{}^{\bd \gd} \bar M_{\bd \gd} $ vanishes, 
the connection $\O_\a{}^{\bd \gd} \bar M_{\bd \gd} $  is flat, 
\begin{subequations}\label{FlatConnection}
\bea
\O_\a{}^{\bd \gd} \bar M_{\bd \gd}  &=& -{\bar g}^{-1}  E_\a \bar g ~, \qquad 
\bar g = \exp \big( \bar L^{\bd \gd} \bar M_{\bd \gd} \big)~,\\
\bar \O_\ad{}^{\b \g}  M_{\b \g}  &=& -{ g}^{-1}  \bar E_\ad  g ~, \qquad 
 g = \exp \big(  L^{\b \g}  M_{\b \g} \big)~.
 \label{FlatConnection.b}
\eea
\end{subequations}

It follows from \eqref{U(1)algebra.a} that the spinor components $E_\a$ of the inverse supervielbein $E_A$ form a closed algebra in the sense that  $\{ E_\a , E_\b \} = C_{\a\b}{}^\g E_\g$. Then the Frobenius theorem implies that $E_\a$ is a linear combination of coordinate supervector fields, 
\begin{subequations} \label{spinor-vielbein}
\bea
E_\a &=& F N_\a{}^\m \hat{E}_\m ~, \quad  \hat{E}_\m = \re^W\pa_\m \re^{-W}   ~, 
\quad W = W^N \pa_N 
~, \label{spinor-vielbein.a}\\
\bar E_\ad &=& \bar F \bar N_\ad{}^{\dot \m} \hat{\bar E}_{\dot \m} ~, \quad  
\hat{\bar E}_{\dot \m} = - \re^{\bar W} \bar \pa_{\dot \m} \re^{-\bar W}   ~,
\quad \bar W = \bar W^N \pa_N~.
\eea
\end{subequations}
Here the matrix $N= (N_\a{}^\m )$ is unimodular, $N \in \sSL(2,{\mathbb C})$, and the scalar $F$ is nowhere vanishing. The complex supervector field $W^N $ is unconstrained.

Consider a covariantly chiral tensor superfield $\J_{\a_1 \dots \a_n}$.
Making use of \eqref{FlatConnection} and \eqref{spinor-vielbein} gives
\bea
\bar \cD_\bd \J_{\a_1 \dots \a_n} =0 \quad \Longleftrightarrow \quad 
\J_{\a_1 \dots \a_n} (x,\q, \bar \q) = g^{-1}
\re^{\bar W} \hat \J_{\a_1 \dots \a_n} (x,\q) ~,
\eea
 where $g$ is given by \eqref{FlatConnection.b}
 
 Local Lorentz transformations correspond to setting  $K^N =0$ in \eqref{K-group}.
 They act only on the matrix $N$ in \eqref{spinor-vielbein.a},   
 \bea 
 N' = \exp \big( \hf K^{ab} \s_{ab} \big) N~,
 \eea
 and therefore it it possible to choose 
 \bea
 N = {\mathbbm 1}~.
 \label{N=1}
 \eea
 This gauge condition is useful for several applications, see below.
 Another useful gauge fixing of the local Lorentz symmetry is 
 \bea
 \O_\a{}^{\bd \gd} = 0~.
 \eea
 
 
 \subsection{The $\L$ gauge group} 
 
 General coordinate transformations correspond to setting  $K^{bc} =0$ in \eqref{K-group}. They act on the building blocks  in \eqref{spinor-vielbein.a} as follows:
 \bea
 F' = \re^K F~,  \quad N' = \re^K N ~, \qquad \re^{W'} = \re^K \re^{W}~.
 \eea
 Once the constrains on the torsion have been partially solved in terms of $F$, $N$ and the complex unconstrained prepotential $W^N$, there may appear an additional gauge freedom. To uncover it, consider a covariantly chiral scalar superfield $\F$
 \bea
 \bar \cD_\ad \F =0 \quad \Longleftrightarrow \quad \F(x,\q,\bar \q) = \re^{\bar W} \hat \F (x, \q) ~, \qquad 
 \bar \pa_{\dot \m} \hat \F =0~.
 \label{chiral5.13}
 \eea
 Its transformation law under \eqref{K-group} is 
 \bea
 \F' = \re^K \F \quad \Longleftrightarrow \quad  \re^{\bar W'} = 
 \re^K \re^{\bar W}  ~, \quad \hat \F' =\hat \F~.
 \eea
 We can introduce a new gauge transformation defined by  
\begin{subequations}\label{Lambda-group}
 \bea
  \re^{\bar W'} &=& 
  \re^{\bar W}  \re^{-\L} ~, \qquad \hat \F' =\re^\L \hat \F = \exp \big(  \l^n \pa_n +\l^\n \pa_\n \big) \hat \F
  ~, \\
  \L &=& \l^N \pa_N = \l^n \pa_n +\l^\n \pa_\n +\l_{\dot \n} \bar \pa^{\dot \n} ~, 
  \quad  \bar \pa_{\dot \m} \l^n = 0~, \quad  \bar \pa_{\dot \m} \l^\n =0~,~~
  \eea
  \end{subequations}
 which does not change $\F$ and which preserves the chirality of $\hat \F$. 
 Of special significance is the fact that the spinor parameter $\l_{\dot \n}$ is unconstrained. 
 It is obvious that the gauge transformations \eqref{Lambda-group} form a group, which is 
  known as the $\L$ gauge group. It turns out that the $\L$-transformation of $W$, 
 \bea
 \re^{W'} = \re^W \re^{-\bar \L} ~, \quad 
 \bar \L = \bar \l^n \pa_n + \bar \l^\n \pa_\n + \bar \l_{\dot \n} \bar \pa^{\dot \n}  ~,
 \eea
 can be supplemented by certain transformations of $F$ and $N$ such that the supervector field $E_\a$, eq. \eqref{spinor-vielbein.a}, does not change. 
In the infinitesimal case, since  
$\d \hat E_\m = - \re^W [\bar \L , \pa_\m ] \re^{-\bar W} $, 
these transformations are:
\bea
\d F= - \hf F \pa_\m \bar \l^\m~, \qquad \d N_\a{}^\m = - N_\a{}^\n \re^W \pa_{(\n} \bar \l^{\m )} ~.
\eea


\subsection{The gravitational superfield} 

Let us analyse the transformation of $W$ under the $K$ and $\L$ gauge groups, 
$ \re^{W'} =  \re^K \re^W \re^{-\bar \L}$. in the infinitesimal case, this reduces to 
\bea
\d W &=& \d W^M \pa_M = K- \bar \L  + O(W) \non \\
&=& (K^m - \bar \l^m) \pa_m + (K^\m - \bar \l^\m) \pa_\m +(K_{\dot \m} - \bar \l_{\dot \m} ) \bar \pa^{\dot \m}
+ O(W)~.
 \eea
 Here the vector parameter $K^m$ is real but otherwise unconstrained, and the spinor parameters $K^\m$ and $\bar \l_{\dot \m}$ are unconstrained. Therefore it is possible to choose a gauge condition 
 \bea
 W = - \ri H~, \qquad H = H^m \pa_m = \bar H~.
 \label{520}
 \eea
 
 A different gauge fixing is possible \cite{GS80}. First one may gauge fix $W$ to have no spinor components, 
 $W= W^n \pa_n$, and then impose the additional condition 
 \bea
 \exp \big( \bar W^n \pa_n \big) x^m = x^m + \ri \cH^m (x,\q, \bar \q) ~, \qquad \bar \cH^m = \cH^m~.
 \label{GravSupGauge}
 \eea
 Given a covariantly chiral superfield \eqref{chiral5.13}, it holds that 
 \bea
  \F(x,\q,\bar \q) = \re^{\bar W} \hat \F (x, \q)   = \hat \F (x +\ri \cH , \q)~.
  \eea
 The residual gauge freedom, which preserves the condition \eqref{GravSupGauge} is determined by considering the variation 
 \bea
 \ri \d \cH^m  &=& \d  \re^{ \bar W } x^m = K  \re^{ \bar W} x^m 
 -  \re^{ \bar W} \L x^m \non \\
 &=& K^m -  \re^{ \bar W} \l^m + \ri K^N \pa_N \cH^m~.
\label{5.24}
 \eea
 Here the right-hand side should be purely imaginary, hence $K^m$ is expressed in terms of $\l^m $ and
 $\bar \l^m$ as follows
 \bea
 K^m = \hf  \re^{\bar W }  \l^m +\hf  \re^W  \bar \l^m
 = \hf \l^m (x +\ri \cH, \q ) + \hf \bar \l^m (x - \ri \cH , \bar \q )~, 
 \label{K-vector}
 \eea
 and the variation $\d \cH^m$ turns into
\bea
\d \cH^m = K^N \pa_N \cH^m +\frac{\ri}{2} \Big( \l^m (x+\ri \cH, \q) - \bar \l^m (x-\ri \cH , \bar \q) \Big)~.
\eea
It is also necessary to require $ \d  \re^{ \bar W }  \q^\m =0$ and $ \d  \re^{ \bar W } \bar \q_{\dot \m} =0$, 
which gives
\bea
K^\m &=& \l^\m (x+\ri \cH, \q) ~, \quad \bar K_{\dot \m} = \bar \l_{\dot \m} (x-\ri \cH, \bar \q)~,
\label{K-spinor}
\eea
 as well as $\L^M = \big( \l^m (x,\q), \l^\m (x,\q), \re^{-\bar W} \re^W \bar \l_{\dot \m} (x,\bar \q) \big)$. 
 Substituting the obtained expressions for $K^\m$, $K^\m $ and $\bar K_{\dot \m}$ into \eqref{5.24}, 
 we arrive at the gauge transformation law of the gravitational superfield, eq. \eqref{H-transformation}.
 
 
 \subsection{Chiral prepotential} 

In order to uncover a remaining prepotential, we first analyse the structure of $R$, $F$ and $E$. These objects are invariant under the local Lorentz transformations, and therefore we can compute them by imposing the gauge condition \eqref{N=1}. In this gauge $E_\a = F \hat E_\a$, $C_{\a\b}{}^\g = 2E_{(\a} \ln F \d_{\b)}{}^\g $
and therefore
the spinor connections are 
\bea
\O_{\a\b\g} = - 2\ve_{\a(\b} E_{\g)} \ln F ~, \qquad 
\bar \O_{\ad\bd\gd} = - 2\ve_{\ad(\bd} \bar E_{\gd)} \ln \bar F ~.
\eea
 Now we can evaluate the relation \eqref{U(1)algebra.a} to end up with explicit expressions for the chiral torsion $R$ and its conjugate $\bar R$:
 \bea
 \bar R  = -\frac 14 \hat E^\m \hat E_\m F^2~, \qquad 
 R = -\frac 14 \hat{\bar E}_{\dot \m} \hat{\bar E}^{\dot \m} \bar F^2~.
 \eea
 Given a scalar superfield $U$, a short calculation gives 
 \bea
 ( \bar \cD^2 - 4 R) U = \hat{\bar E}_{\dot \m} \hat{\bar E}^{\dot \m} ( \bar F^2 U)~.
 \label{529}
 \eea
  
In order to compute $E^{-1} = {\rm Ber} (E_A{}^M)$, we introduce a semi-covariant frame 
\bea
\hat E_A = (\hat E_a , \hat E_\a , \hat{\bar E}^\ad ) = \hat E_A{}^M \pa_m ~, 
\qquad \hat E_a = -\frac{\ri}{4} ({\tilde \s}_a)^{\ad \a} \{ \hat E_\a  , \hat{\bar E}_\ad \}~,
\eea
which is constructed in terms of the prepotential 
$W^M $ and its conjugate, in accordance with \eqref{spinor-vielbein}.
 In the gauge \eqref{N=1}, the inverse supervielbein $E_A$ is related to $\hat E_A$ as follows:
\begin{subequations}
\bea
E_\a &=& F \hat E_\a ~, \qquad {\bar E}^\ad = \bar F \hat{\bar E}^\ad ~, \\
E_a &=& F \bar F \hat E_a + \frac{\ri}{4} F ({\tilde \s}_a)^{\ad \a} \big( \O_{\ad\a}{}^\b 
- \d_\a{}^\b \bar E_\ad \ln F \big) \hat E_\b \non \\
&& \phantom{F \bar F \hat E_a} + \frac{\ri}{4} \bar F ({\tilde \s}_a)^{\ad \a} 
\big( \bar \O_{\a\ad}{}^\bd 
- \d_\ad{}^\bd  E_\a \ln \bar F \big) \hat{\bar E}_\bd~. 
\eea
\end{subequations}
It follows that
\bea
E^{-1} = F^2 \bar F^2 \hat E^{-1}~, \qquad \hat E^{-1} = {\rm Ber} (\hat E_A{}^M) ~.
\label{532}
\eea

So far, $F$ appears to be unconstrained. However, it follows from the algebra of covariant derivatives 
that 
\bea
(-1)^{\ve_B} \cT_{\a B }{}^B =0~, 
\eea
while the direct evaluation of this structure gives (see, e.g., \cite{BK} for the technical details)
\bea
-(-1)^{\ve_B} \cT_{\a B }{}^B = E_\a \ln \big[E^{-1}
F^2 (1 \cdot
\re^{ \stackrel{\leftarrow}{W} }) \big] = E_\a \ln \big[\hat E^{-1}
\bar F^2F^4 (1 \cdot
\re^{ \stackrel{\leftarrow}{W} }) \big]~, 
\eea 
where the operator $\stackrel{\leftarrow}{W} $ is defined by 
\bea
U \stackrel{\leftarrow}{W} 
= (-1)^{\ve_M} \pa_M (U W^M ) \quad \Longrightarrow \quad 
(U \cdot \re^{ \stackrel{\leftarrow}{W} } )=  (1 \cdot \re^{ \stackrel{\leftarrow}{W} } ) \re^W U~.
\eea
We conclude that $\bar \vf^{-3}:=\bar F^2F^4 \hat E^{-1} (1 \cdot
\re^{ \stackrel{\leftarrow}{W} }) $ is annihilated by $E_\a$. This result can be  equivalently written as
\bea
 \vf^{-3} =F^2 \bar F^4  \hat E^{-1} (1 \cdot
\re^{ \stackrel{\leftarrow}{W} }) ~, \qquad \bar E_\ad \vf =0~.
\label{536}
\eea
By construction, the chiral superfield $\vf$ is nowhere vanishing. 
It follows that 
\bea
F = \vf^{1/2} \bar \vf^{-1} (1 \cdot
\re^{ \stackrel{\leftarrow}{W} })^{- 1/3}(1 \cdot
\re^{ \stackrel{\leftarrow}{\bar W} })^{1/6} \hat E^{1/6}~,
\eea
and then eq. \eqref{532} gives
\bea
E = \bar \vf \vf \big[ \hat E (1 \cdot
\re^{ \stackrel{\leftarrow}{W} }) (1 \cdot
\re^{ \stackrel{\leftarrow}{\bar W} })\big]^{1/3} ~.
\eea

The covariantly chiral superfield $\vf$ is called the chiral prepotential. It turns out that, modulo purely gauge degrees of freedom, the covariant derivatives are expressed in terms of $W^M$, $\vf$ and their conjugates. 
These are the prepotentials for the GWZ superspace geometry. The transformation law of $\vf$ follows from \eqref{536} 
\bea
\d \vf^3 = K^M \pa_M \vf^3 + \vf^3 \re^{\bar W} (\pa_m \l^m - \pa_\m \l^\m) ~.
\label{539}
\eea
If we represent the chiral prepotential in the form 
\bea 
\vf =  \re^{\bar W} \hat \vf ~, \qquad 
 \bar \pa_{\dot \m} \hat \vf =0~,
 \eea
then the transformation law \eqref{539} will turn into 
\bea
\d \hat \vf^3 = \l^M \pa_M \hat \vf^3 + \hat \vf^3 (\pa_m \l^m - \pa_\m \l^\m)  
= \pa_m( \l^m \hat \vf^3 ) - \pa_\m ( \l^\m \hat \vf^3) ~.
\label{chiral_density}
\eea
This is the transformation law of a chiral density.

By construction, the prepotential $W^M $ is invariant under the super-Weyl transformations 
\eqref{superweyl}. It is a short calculation to see that the chiral prepotential $\vf$ transforms  as 
\bea
\d_\s \vf = - \s \vf~.
\eea
In conformal supergravity, the super-Weyl transformations belong to the gauge group. Making use of the super-Weyl gauge freedom allows one to impose the gauge $\vf=1$. Therefore, the gravitational superfield is the only prepotential in conformal supergravity, modulo purely gauge degrees of freedom.


\subsection{Chiral action}\label{section5.5}

It follows from the above analysis that $E$ can be written as
$E= \vf^3 \bar F^2 (1 \cdot\re^{ \stackrel{\leftarrow}{\bar W} })$. In conjunction with the identity 
\eqref{529}, the chiral action 
\eqref{3.30} can be rewritten as follows 
\bea
S_{\rm c} = -4  \int\text{d}^4x \rd^2 \q \rd^2 \bar \q \, \vf^3 (1 \cdot\re^{ \stackrel{\leftarrow}{\bar W} })
\mathcal{L}_{\rm c}\,
\frac{\bar F^2 \U}{\hat{\bar E}_{\dot \m} \hat{\bar E}^{\dot \m} ( \bar F^2 \U)
} ~,
\label{542}
\eea
We now recall the well-known result for a change of variable in superspace \cite{GS80} (see \cite{BK} for a pedagogical derivation). Given a first-order differential operator $K=K^N \pa_N$, it holds that
\bea
 z'{}^M = \re^K z^M \quad \Longrightarrow \quad 
 {\rm Ber} (\pa_M z'^N ) = (1 \cdot\re^{ \stackrel{\leftarrow}{K} })~,
 \label{543}
 \eea
and therefore $\int \rd z' L(z') = \int \rd z (1 \cdot\re^{ \stackrel{\leftarrow}{K} }) \re^K L(z)$.
As follows from \eqref{chiral5.13}, the covariantly chiral superfields depend on 
chiral variables $\tilde{x}^m $ and $\tilde{\q}^\m$, 
\bea
\hat{\bar E}_{\dot \m} \F =0  \quad \Longrightarrow \quad 
 \F(z) =  \hat \F (\tilde{x}, \tilde{\q}) ~, \qquad 
\tilde{z}^M =  (\tilde{x}^m, \tilde{\q}^\m , \tilde{\bar \q}_{\dot \m} ) = \re^{\bar W} z^M~.
\eea
In the variables $\tilde{z}^M$, the operator $\hat{\bar E}^{\dot \m} $
becomes a partial derivative, $\hat{\bar E}^{\dot \m}  =  \pa / \pa \tilde{\bar \q}_{\dot \m} $.
Now, making use of \eqref{543} in \eqref{542} leads to the following simple result
\bea
S_{\rm c} =   \int\text{d}^4x \rd^2 \q  \, \hat{\vf}^3 
\hat{\mathcal{L}}_{\rm c}~.
\eea
Here we have denoted  $\tilde{x}^m$ and $\tilde{\q}^\m $ simply as $x^m $ and $\q^\m$.
This result shows that the chiral integration measure  in \eqref{chiralAc} is 
\bea
\cE = \vf^3~,
\eea
and this interpretation agrees with the transformation law \eqref{chiral_density}.


\section{Matter multiplets in conformal supergravity}

In this section we introduce the most popular matter multiplets 
and describe several famous models for them. Practically all results will be presented in conformal superspace. They can be recast in terms of the $\sU(1)$ or GWZ superspace geometries by making use of the degauging formalism
described in section \ref{Section4}.


\subsection{Scalar multiplet}
\label{subsection6.1}

The minimal scalar multiplet was introduced by Wess and Zumino in their first paper on supersymmetry \cite{WZ74}. In conformal superspace, minimal scalar multiplets are described in terms of covariantly chiral primary scalar superfields. Such a superfield 
$\f$ obeys the constraints
$K^B \f =0$ and ${\bar \de}^\ad \f=0$. In general, every covariantly chiral primary superfield $\f$ of definite dimension $\D$ satisfies equation \eqref{chirals}.
If we do not assume $\f$ to be an eigenvector of $\mathbb D$ then it must hold that 
\bea
K^B \f =0~, \quad  {\bar \de}^\bd \f=0\quad \Longrightarrow \quad
   \mathbb{Y} \f =  -\frac 23  {\mathbb D} \f~.  
\label{cov-chiral-1}
\eea
The superfield $\f$ contains three independent component fields which can be chosen as follows: $\vf:= \f|$, $\eta_\a:=\nabla_\a \f |$ 
and $F:=- \frac 14 \nabla^2 \f |$. In theories with at most two derivatives at the component level, the complex scalar $F$ is an auxiliary field.

As a simple example of a supergravity-matter system, we consider a  curved superspace extension of the massless Wess-Zumino model \cite{WZ74}. It corresponds to choosing a canonical dimension for the chiral scalar, 
${\mathbb D} \f = \f$. The action is
\bea
S_{\rm WZ} = \int\text{d}^{4|4}z \, E \, \bar \f \f 
 +\Big\{ \frac{\l}{3} \int \rd^4x\rd^2\q\, \cE \, \f^3 +{\rm c.c.} \Big\} ~,
\eea
 with $\l$ a complex coupling constant.
 

\subsection{Superconformal sigma model} 

In Minkowski superspace ${\mathbb M}^{4|4}$,  general $\cN=1$ supersymmetric two-derivative theories of scalar multiplets are nonlinear $\s$-models which are
described by chiral scalar superfields $\f^I$ and their conjugates $\bar \f^{\bar I}$ taking their values in an arbitrary K\"ahler manifold $\cM$ \cite{Zumino79}.
 In supergravity, however, $\s$-model couplings turn out to be more restrictive, 
and the target space $\cM$ must be a K\"ahler-Hodge manifold
\cite{Witten:1982hu}. Within the locally superconformal setting, this means that we have to consider a superconformal sigma model on 
a K\"ahler cone (see, e,.g.,  \cite{FVP,GR} for a more detailed discussion). 
Here, our goal is to show how these restrictions emerge.  

Let $N(\f, \bar \f)$ be the 
K\"ahler potential of 
$\cM$, and  $g_{I\bar J} = \pa_I \pa_{\bar J} N \equiv N_{I \bar J}$ its K\"ahler metric.
 We start with a naive curved-superspace extension of the $\cN=1$ 
supersymmetric  $\s$-model  action\footnote{An overall minus sign is inserted in \eqref{6.5} in order to give the correct sign for the Einstein-Hilbert term at the component level, if \eqref{6.5} is viewed as the supergravity-matter action, see \cite{FVP} for more details.}  
\bea
S= -\int {\rm d}^{4|4}z\, E\,N(\f, {\bar \f})~, \qquad 
K^B \f^I=0~, \quad\deb^\bd \f^I=0
~.
\label{6.5}
\eea
Here the dynamical variables $\f^I$ are postulated to be covariantly chiral primary scalar superfields. Since $\f^I$ are local holomorphic coordinates, 
${\mathbb D} \f^I$ and ${\mathbb Y} \f^I$ must be holomorphic vector fields on $\cM$, 
 \bea
{\mathbb D}\f^I =  \c^I (\f) \quad  \Longleftrightarrow \quad {\mathbb Y}\f^I =  -\frac{2}{3}\c^I (\f)~,
\label{genN=2sctr}
\eea
where we have used \eqref{cov-chiral-1}. 
The action (\ref{6.5}) must be locally superconformal. 
Then, in accordance with \eqref{3.27},
$N$ must be neutral under the $\sU(1)_R$ group and have dimension $+2$,
and therefore
\begin{subequations}
\bea
\c^I (\f) \pa_I N (\f , \bar \f ) &=&\bar \c^{\bar I} (\bar \f)  \bar \pa_{\bar I} N (\f , \bar \f ) ~, \\
\c^I (\f) \pa_I N (\f , \bar \f ) 
&=& 
\bar \c^{\bar I} (\bar \f) \bar \pa_{\bar I} N(\f , \bar \f ) = N(\f , \bar \f) ~.
\label{6.7}
\eea
\end{subequations}
Differentiating the condition $\c^I  \pa_I N =N$ with respect to $\bar \pa_{\bar J}$ gives
\bea
\c^I (\f) g_{I\bar J} (\f , \bar \f ) = \bar \pa_{\bar J} N (\f , \bar \f) = \bar \c_{\bar J} (\f, \bar \f) 
\quad \implies \quad 
\c^I (\f) =g^{I\bar J}  \bar \pa_{\bar J} N~.
\label{6.8}
\eea
The obtained relations have several nontrivial implications. First of all, the equations 
\eqref{6.7} and \eqref{6.8} imply that $N$ is a globally defined function on $\cM$, 
\bea
 N =g_{I\bar J} \c^I  {\bar \c}^{\bar J}~.
 \label{7.9}
 \eea
Therefore,  the K\"ahler two-form, 
 $ \O=2\ri \,g_{I \bar J} \, \rd \f^I \wedge \rd \bar \f^{\bar J}$,  
is exact, hence $\cM$ is necessarily non-compact. 
Secondly, it follows that $\chi^I$ is a homothetic conformal Killing vector field
\bea
\nabla_I \c^J = \d_I^J~, \qquad {\bar \nabla}_{\bar I} \c^J ={\bar \pa}_{\bar I} \c^J =0~.
\eea

The sigma model (\ref{6.5}) can be generalised  to include a superpotential
\bea
S= -\int {\rm d}^{4|4}z\,E\, 
 N(\f, {\bar \f})
 + \Big\{\int 
  \rd^4x\rd^2\q\, \cE \,
 W(\f)   +{\rm c.c.}
\Big\}
~.
\label{4.33}
\eea
Here $W(\f)$ is a holomorphic scalar field on the target space. It should obey 
${\mathbb D}W(\f)=3W(\f)$ and ${\mathbb Y}W(\f)=-2W(\f)$, which imply 
the homogeneity condition
\bea
\c^I (\f) \pa_I W (\f ) = 3 W(\f )
~.
\eea

It should be mentioned that local complex coordinates in $\cM$ can be chosen in such a way that $\c^I (\f) =\f^I$.
Then the K\"ahler potential $N(\f, {\bar \f}) $ obeys the following homogeneity condition:
\bea
\f^I  \pa_I N(\f, \bar \f) =  N( \f,   \bar \f)~.
\label{Kkahler2}
\eea


\subsection{Superconformal higher-derivative sigma model}

Here we discuss a higher-derivative superconformal $\s$-model which was originally introduced in the GWZ superspace as an induced action \cite{Kuzenko:2020jzb}. It appears that its uplift to conformal superspace cannot be given solely in terms of the conformally covariant derivatives $\nabla_A$ and should explicitly involve connection superfields. 

 Let $K(\f^I, \bar \f^{\bar J})$ be the 
K\"ahler potential of an {\it arbitrary}  K\"ahler manifold $\cM$.  We introduce a higher-derivative 
locally supersymmetric theory described in terms of covariantly chiral scalar superfields
$\f^I$, $\bar \cD^\ad \f^I=0$, which are neutral under the super-Weyl transformations, 
$\d_\s \f^I =0$. The higher-derivative action proposed in \cite{Kuzenko:2020jzb} is 
\bea
S &=& \frac{1}{16}  \int 
\rd^{4|4}z \, 
E \, g_{I \bar J} (\f, \bar \f) 
\Big\{ \mathfrak{D}^2 \f^I \bar{\mathfrak{D}}^2 \bar \f^{\bar J} 
-8 G_{\a\ad}\cD^\a \f^I  \bar \cD^\ad \bar \f^{\bar J} \Big\}\non \\
&&+ \frac{1}{16} \int 
\rd^{4|4}z
 \, E\,
\Big\{ \a
R_{I \bar J K  \bar L} (\f, \bar \f) +\b g_{I \bar J} (\f, \bar \f)
g_{K  \bar L} (\f, \bar \f) \Big\} \non \\
&&\qquad \qquad \times 
 \cD^\a \f^I \cD_\a \f^K \bar \cD_\ad 
\bar \f^{\bar J} \bar \cD^\ad \bar \f^{\bar L}~,
\label{SCHD}
\eea
where $g_{I\bar J} = \pa_I \pa_{\bar J} K$ is the K\"ahler metric, 
$R_{I \bar J K  \bar L} $ 
the Riemann curvature of the K\"ahler manifold, 
and $\mathfrak{D}^2 \f^I $ is defined as follows\footnote{The operator $\mathfrak D^2$ in \eqref{nabla2} should not be confused with ${\mathfrak D}^\a {\mathfrak D}_\a$ in $\sU(1)$ superspace.}
\bea
\mathfrak{D}^2 \f^I := \cD^2 \f^I + \G^I_{KL} \cD^\a \f^K \cD_\a \f^L~.
\label{nabla2}
\eea
We recall that the Christoffel symbols $\G^I_{KL} $ and the curvature
$R_{I \bar J K  \bar L} $ are given by the expressions 
\bea
\G^I_{JK} = g^{I \bar L} \pa_J \pa_K \pa_{\bar L} K ~,
\quad R_{I \bar J K  \bar L} = \pa_I \pa_K \pa_{\bar J} \pa_{\bar L}  K 
-g^{M \bar N} \pa_I \pa_K \pa_{\bar N} K  \pa_{\bar J} \pa_{\bar L}  \pa_M K ~.
\eea
It is an instructive exercise to show that the action \eqref{SCHD} is super-Weyl invariant. 
This action is manifestly invariant under K\"ahler transformations
\bea
K(\f, \bar \f) \to K(\f, \bar \f) + \L(\f) + \bar \L (\bar \f) ~,
\label{Kahler}
\eea
with $\L(\f) $ being an arbitrary holomorphic function.

The super-Weyl invariance of  \eqref{SCHD} may be traced  to the existence of a superconformal
operator $ \D$ introduced in \cite{BdeWKL}.  In conformal superspace, this operator is defined to act on a primary chiral weight-zero scalar $\bar \f$ as 
\begin{subequations}\label{superFT}
\bea
\D \bar \f = -\frac{1}{64} \bar \nabla^2  \nabla^2 \bar \nabla^2 \bar \f
\eea
and the resulting weight-three chiral superfield is primary, 
\begin{align}
K^B \bar  \f =0~, \quad  {\nabla}^\b \bar \f=0,\quad {\mathbb D} \bar \f =0 \quad \implies \quad
K^B  \D \bar \f =0~,\quad \bar \nabla^\bd  \D \bar \f =0~.~~
\end{align} 
\end{subequations}
Degauging $ \D\bar \f $ to the GWZ superspace gives
\bea
 \D\bar \f := -\frac{1}{64} (\bar \cD^2 -4R ) \Big\{ \cD^2 \bar \cD^2 \bar \f
+ 8 \cD^\a (G_{\a\ad}\bar \cD^\ad \bar \f)\Big\}~.
\label{Delta22}
\eea
The  super-Weyl transformation law of this superfield is
$\d_\s \D\bar \f = 3\s \D\bar \f $.
For any covariantly chiral scalars $\f$ and $\j$, it holds that 
\bea
\int \rd^4 x \,\rd^2 \q \,\cE \, \j \D \bar \f 
= \frac{1}{16}\int \rd^{4|4}z \,E \,
 \Big\{ (\cD^2 \j) \bar \cD^2 \bar \f 
-
8 (\cD^\a \j) G_{\a\ad}\bar \cD^\ad \bar \f\Big\} ~.~~~~
\label{symmetric}
\eea
If the chiral scalars $\f$ and $\j$
are inert under the super-Weyl transformations, this functional 
is super-Weyl invariant.

The operator \eqref{superFT} is a supersymmetric generalisation of the  conformal fourth-order scalar operator 
in curved space
\bea \label{D_0}
\Delta_0 = \Box \Box -  \nabla^a \big(
	2 \mathcal{R}_{ab} \nabla^b 
	- \tfrac{2}{3} \mathcal{R} \nabla_a
	\big)~, \qquad \Box = \nabla^a \nabla_a
\eea
discovered by Fradkin and Tseytlin \cite{FT1982}. Here $\nabla_a$ denotes the torsion-free Lorentz covariant derivative, with $\mathcal{R}_{ab}$ and $ \mathcal{R} $ being its Ricci tensor and scalar curvature, respectively. The operator \eqref{superFT} was constructed for the first time in \cite{BdeWKL} using the conformal superspace approach, although there had been earlier attempts to construct such an operator, see the discussion in \cite{Butter:2013ura}.
Making use of its degauged form, eq.  \eqref{Delta22}, a new representation for the nonlocal action generating 
the super-Weyl anomalies was derived in  \cite{Butter:2013ura}.


\subsection{Tensor multipet} 

The massless tensor multiplet was introduced by Siegel \cite{Siegel-tensor} as a dual version of the minimal scalar multiplet. In conformal superspace, it is described by a primary covariantly chiral spinor superfield $\eta_\a$ of dimension 3/2, 
\bea
K^B \eta_{\a} =0~, \qquad \bar \nabla^\bd \eta_{\a} =0 \qquad {\mathbb D} \eta_{\a} = \frac 32 \eta_{\a} ~,
\label{6.19}
\eea
which is defined modulo gauge transformations
\bea
\d \eta_\a = - \frac{\ri}{4} \bar \nabla^2 \nabla_\a U ~, \qquad \bar U =U~,
\eea
with the gauge parameter being a primary dimensionless real scalar. 
The descendant 
\bea
\mathbb  G = \hf \big( \nabla^\a \eta_\a + \bar \nabla_\ad \bar \eta^\ad \big) = \bar{\mathbb G} 
\label{G}
\eea
is a gauge-invariant field strength. It has the following properties:
\bea
K^B \mathbb G =0 ~, \qquad
\bar \nabla^2 \mathbb G =0 ~,
\qquad {\mathbb D} \mathbb G = 2\mathbb G~.
\label{6.22}
\eea
These constraints define a {\it real linear multiplet}. Such superfields were originally introduced by Ferrara, Wess and Zumino \cite{FWZ} to describe flavour current multiplets.

The superconformal tensor multiplet is described by the action \cite{deWR}
\bea
S = - \int {\rm d}^{4|4}z \,E\,  \mathbb  G \ln \frac{\mathbb G}{\bar \f \f}  
~,\qquad K^B \f =0~, \quad \bar \nabla^\bd \f =0~, \quad {\mathbb D} \f = \f ~.
\label{ImprovedTensor}
\eea
Both $\mathbb G$ and $\f$ are assumed to be nowhere vanishing. Dependence of the action 
\eqref{ImprovedTensor} on $\f$ and $\bar \f$ is fictitious, since the action remains unchanged 
under transformations $\f \to \re^\s \f$,  where $\s$ is an arbitrary covariantly chiral weight-zero scalar. In the literature, \eqref{ImprovedTensor} is referred to as the  model for an improved tensor multiplet \cite{deWR}. It is a unique superconformal representative in the family  of tensor multiplet models introduced in \cite{Siegel-tensor}.


\subsection{Three-form multiplet} 

Let us consider the representation \eqref{N=2chiralPsi} for $n=0$.
The unconstrained prepotential $\j$ in \eqref{chiral-prepotential} is necessarily complex for $\D \neq 3$.  In the $\D =3$ case, however, one can impose the reality condition $\bar \j = \j = P$.
This leads to the three-form multiplet\footnote{In global supersymmetry, the three-form multiplet was originally proposed by Gates \cite{Gates}.}
 described by the primary covariantly chiral scalar 
\bea
\P = - \frac 14 \bar \nabla^2 P~, \quad \bar P  = P ~, \quad K^B P =0~, \quad {\mathbb D} P = 2P~.
\label{6.24}
\eea
The main difference of the three-form multiplet from the minimal scalar multiplet is that 
the imaginary part of the auxiliary field $F:=-\frac 14 \nabla^2 \P|$ of $\P$
is the field strength of a gauge three-form. 

The prepotential $P$ in \eqref{6.24} is defined modulo gauge transformations 
$\d P = \mathbb G$, where the gauge parameter is a real linear superfield, eq.  \eqref{G}. 
The simplest superconformal and gauge-invariant action to describe the dynamics of this multiplet is given by 
\bea
S &=& \int\text{d}^{4|4}z \, E \, \Big\{ \big(\bar \P \P\big)^{1/3} +  2\kappa P \Big\} \non \\
&=&\int\text{d}^{4|4}z \, E \, \big(\bar \P \P\big)^{1/3} 
+ \Big\{ \k\int 
   \rd^4x\rd^2\q\, \cE \,
\P   +{\rm c.c.}
\Big\}
~,
\eea 
where $\k$ is a real coupling constant.


\subsection{Non-minimal scalar multiplet}

We next turn to a non-minimal scalar multiplet.\footnote{In global supersymmetry, it was introduced by Gates and Siegel \cite{GS81}.}
In conformal superspace it is described 
by a primary complex scalar superfield $\G$ satisfying the constraint
\begin{subequations}\label{complex-linear}
\bea 
K^B\G=0~,\quad
\deb^2\G=0
 \quad  \implies \quad \mathbb{Y} \G= \frac 23 \big(2 - {\mathbb D}\big) \G~.
\eea
We choose to parametrise the dimension of $\G$ as 
\bea
{\mathbb D}\G=\frac{2}{3n+1}\G
\quad \implies \quad
\mathbb{Y} \G=\frac{4n}{3n+1}\G~,
\eea
\end{subequations}
following the notation introduced in \cite{SG}. For $n\neq 0, -1/3$, the constraint \eqref{complex-linear} defines a {\it complex linear superfield}. In the $n=0$ case, the $\sU(1)_R$ charge of $\G$ is equal to zero, and $\G$ can be subject to the reality condition $\bar \G = \G$, 
which corresponds to the real linear multiplet \eqref{6.22}.
The general solution to the constraint \eqref{complex-linear} is 
\bea
\G = \bar \nabla_\ad \bar \J^\ad ~,
\eea
where the unconstrained prepotential $ \bar \J^\ad $ may be chosen to be primary. It is defined  
modulo gauge transformations $\d  \bar \J^\ad = \bar \nabla_\bd  \bar \l^{(\ad \bd)}$, where the gauge parameter may be chosen to be primary.

A unique superconformal model, which is constructed solely in term of $\G$ and $\bar \G$ and involves at most two derivatives at the component level, is given by  
\bea
S=- \frac{1}{n}\int {\rm d}^{4|4}z\,E\, \big( \bar \G \G\big)^{(3n+1)/2}~.
\label{non-min6.29}
\eea

In global supersymmetry, it was observed by Deo and Gates \cite{DG85} that the complex linear 
constraint $\bar D^2 \G =0$ admits a deformation $ \bar D^2 \G = -4 \X$
 in the presence of a chiral scalar $\X$. This idea is compatible 
with local superconformal symmetry. Indeed, in conformal superspace the constraint 
\eqref{complex-linear} can be deformed 
  to define the following \emph{improved} linear constraint
\begin{align}
-\frac{1}{4}\deb^2\U = \X~,\qquad K^B \X=0~, \quad \bar \nabla^\bd \X=0~, \quad
{\mathbb D}\X=\frac{3(n+1)}{3n+1}\X~.
\label{NM6}
\end{align}
In general, $\X$ may be a function of matter chiral scalars, $\X = \X(\f)$,
see \cite{DG85,TartaglinoMazzucchelli:2004vt}. 
Such constraints naturally arise in the framework of  the $\cN=1$ superfield description of $\cN=2$ supersymmetric sigma models \cite{Kuzenko:2006nw}.

It follows from \eqref{NM6} that  the choice $n=-1$ is special in the sense that $\X$ becomes dimensionless, and therefore one can impose the superconformal constraint 
\begin{align}
K^B \U =0~, \quad -\frac{1}{4}\deb^2\U = \m = {\rm const} \quad \implies \quad 
{\mathbb D} \U = -\U~.
\label{AdScompensator}
\end{align}
This multiplet originates as the compensator of the non-minimal AdS supergravity proposed in \cite{BK11}. It is also used to describe the dynamics of a Goldstino \cite{KTyler}.


\subsection{Vector multiplet}

The Abelian vector multiplet was introduced by Wess and Zumino in their first paper on supersymmetry \cite{WZ74}. Its Yang-Mills extension was derived by Ferrara and Zumino \cite{Ferrara:1974pu} and, independently, by Salam and Strathdee \cite{Salam:1974ig}.
Here we briefly review the conformal superspace formulation for the Abelian vector multiplet and related superconformal models.

The Abelian vector multiplet is described by a scalar dimension-zero prepotential $V$ defined modulo gauge transformations of the form 
\bea
\d_\L V = \L + \bar \L ~, \qquad \bar \nabla_\ad \L =0~.
\eea
Both the prepotential $V$ and the chiral gauge parameter may be chosen to be primary. Associated with $V$ is the primary chiral spinor descendant 
\bea
W_\a = - \frac 14 \bar \nabla^2 \nabla_\a V ~, \qquad K^B W_\a = 0~, \quad \bar \nabla^\bd W_\a =0~, 
\quad {\mathbb D} W_\a = \frac 32 W_\a~,
\eea
which is gauge invariant, $\d_\L W_\a =0$. 
The field strength $W_\a$ is a reduced chiral superfield in the sense that it obeys the 
reality condition $\nabla^\a W_\a =\bar \nabla_\ad \bar W^\ad \equiv \nabla W$.
It should be pointed out that $\nabla W$ is a primary dimension-2 superfield. 
Modulo purely gauge degrees of freedom, 
the independent components of $V$ can be chosen as follows: $\eta_\a = W_\a| $, $v_{\a\ad} =\hf  [\nabla_\a, \bar \nabla_\ad ] V| $ and  $D= -\hf \nabla W|$.

Dynamics of the free vector multiplet is described by the   action \cite{WZ}
\bea
S =
\frac{1}{4}  \int \rd^4 x \rd^2 \q   \, \cE\,
W^2 +{\rm c.c.} 
\label{superMaxwell}
\eea
 For a single vector multiplet
this is a unique locally superconformal action with at most two derivatives at the component level. In the case that $(\nabla W)^{-1}$ exists, nonlinear superconformal actions exist of the form \cite{K19}
\bea
S &=&
\frac{1}{4}  \int \rd^4 x \rd^2 \q   \, \cE\,
W^2 +{\rm c.c.}
+ \frac 14 \int \rd^{4|4} z  \, E\,
\frac{W^2\,{\bar W}^2}{(\nabla W)^2}\,
{\mathfrak H} (u , \bar u)~,
\label{superMaxwell-conf}
\eea
where $u  := \frac{1}{8}   \nabla^2 \big[ W^2 (\nabla W)^{-2}\big]$ is a primary dimensionless antichiral superfield, and 
${\mathfrak H}(z,\bar z)$ is a real function of a complex variable.
This family includes a unique $\sU(1)$ duality-invariant theory \cite{BLST,K21}
\bea
S &=&
\frac{1}{4} \cosh \g \int  \rd^4 x \rd^2 \q  \,\cE \, W^2 +{\rm c.c.}
+ \frac{1}{4}\sinh \g   \int \rd^{4|4} z  \,E \,
\frac{W^2\,{\bar W}^2}{(\nabla W)^2\sqrt{u\bar u} }~,
\label{ModMax}
\eea
where the coupling constant $\g$ must be non-negative \cite{BLST}.\footnote{The general formalism for $\sU(1)$ duality-invariant $\cN=1$ and $\cN=2$ supersymmetric theories was developed in \cite{KT1}.} 
This nonlinear extension of the supersymmetric Maxwell action \eqref{superMaxwell}
 is called  the super ModMax theory.

Within the GWZ superspace formalism, the action \eqref{ModMax} can be rewritten in a simpler, 
albeit not manifestly superconformal form, originally given in  \cite{BLST,K21}
\bea
S &=& \frac{1}{4} \cosh \g \int  \rd^4 x \rd^2 \q  \,\cE \, W^2 +{\rm c.c.}
+ \frac{1}{4}\sinh \g   \int \rd^{4|4} z \,E \,
\frac{W^2\,{\bar W}^2}{ \sqrt{ {\bf u} \bar {\bf u}} }~,
\eea
where ${\bf u}  := \frac{1}{8} \cD^2  W^2$. In order to make direct contact with the 
$\sU(1)$ duality-invariant formalism of \cite{KT1}, this action can be rewritten in the form 
\begin{subequations}
\bea
S &=&
\frac{1}{4} \int  \rd^4 x \rd^2 \q  \,\cE \, W^2 +{\rm c.c.}
+ \frac14  \int \rd^{4|4} z \,E \,
W^2\,{\bar W}^2\,
\L\left({\bf u},
{\bar {\bf u}}\right)~, \\
&& \L({\bf u}, \bar {\bf u}) = \frac{\sinh \g}{\sqrt{{\bf u}\bar {\bf u}} } +\hf (1-\cosh \g) 
\Big( \frac{1}{\bf u} +\frac{1}{\bar {\bf u}} \Big)~.
\eea
\end{subequations}

A large class of other interesting, and not necessarily superconformal, models for supersymmetric nonlinear electrodynamics are based on deforming  the super-Maxwell action \eqref{superMaxwell}  by a self-interaction  $\int \rd^{4|4} z  \, E\, \cL$, where 
\begin{subequations}
\bea
\cL&=&W^2\bar{W}^2 \cH\big(\o,{\bar{\o}},\de W,\mathfrak C\big)~, \qquad 
\o :=  \frac{1}{8}   \nabla^2 \big[ {W^2}{\mathfrak C}^{-2} \big]~,  \\
 K^B {\mathfrak C} &=&0 ~, \quad
 {\mathbb D} {\mathfrak C} = 2{\mathfrak C}~, \quad \bar {\mathfrak C} = {\mathfrak C}~.
\eea 
\end{subequations}
Here $\mathfrak C$ is a conformal compensator associated to an off-shell Poincar\'e supergravity (see the next section), while the composite 
$\cH$ is constrained to be a real primary superfield of  dimension $-4$.
  Well-known theories of this type are, for instance, the supersymmetric Born-Infeld theory \cite{CF}, and the  generalised Fayet-Iliopoulos terms in supergravity without gauged $R$-symmetry, which were  recently introduced in \cite{Cribiori:2017laj}.

The supersymmetric Yang-Mills multiplet is well reviewed in the literature, see, e.g., \cite{GGRS,WB,FVP,BK}, and its description in conformal superspace does not bring in 
new features. We refer the interested reader to the literature, see \cite{Kugo:2016lum}.


\section{Off-shell models for pure supergravity}

As discussed in the introduction, there are several  off-shell formulations for pure supergravity, including the old minimal \cite{Siegel77-77,WZ,old1,old2}, new minimal 
\cite{SohniusW1,SohniusW3} and non-minimal \cite{Breitenlohner,Siegel77-80,SG} theories. Here we present their formulations in conformal superspace.
Due to space limitations, a discussion of general supergravity-matter systems is beyond the scope of this review.

As discussed in section \ref{N1_conformal-superspace}, conformal superspace can be identified with a pair $(\cM^{4|4}, \nabla)$. In the case of Poincar\'e or AdS supergravity, the superspace geometric setup is a triple $(\cM^{4|4}, \nabla, {\mathfrak C})$, where $\mathfrak C$ is a compensator. The latter is a primary constrained scalar superfield 
such that (i) $\mathfrak C$ is nowhere vanishing (more precisely, ${\mathfrak C}^{-1}$ exists); and (ii) the dimension of $\mathfrak C$ is non-zero. These conditions imply that the local scale 
 and local $\sU(1)_R$ gauge freedom can be used to impose the gauge condition $\mathfrak C =1$. If the compensator is real, $\bar{\mathfrak C}=\mathfrak C$, the required gauge condition is achieved by applying a local scale transformation.  


\subsection{Old minimal supergravity}
\setcounter{equation}{0}
\label{old-minimal}

In the old minimal formulation for supergravity, the compensator is a nowhere vanishing primary chiral scalar $\f$, eq. \eqref{chirals}, of non-zero dimension $\D$. Since $\f^{-1} $ exists, 
the primary chiral scalar $\F=\f^{1/\D} $ is also nowhere vanishing and its dimension is canonical, ${\mathbb D} \F = \F$. It is $\F$ and its conjugate $\bar \F$ which are  chosen as the compensators in old minimal supergravity. 

The action functional for pure old minimal supergravity is given by
\bea
S_{\text{old-minimal}}=-3\int {\rm d}^{4|4}z
\,E\,\bar \F \F
 +\Big\{ \m \int \rd^4x\rd^2\q\, \cE \, \F^3 +{\rm c.c.} \Big\} ~.
\label{old-minimal-action}
\eea
The choice $\m=0$ corresponds to Poincar\'e supergravity. For $\m\neq 0$ the action describes AdS supergravity. Let us analyse the equations of motion for this theory. The chirality constraint on $\F$ and its equation of motion can be written as 
\begin{subequations}\label{EoM1}
\bea
\bar \nabla_\ad \F^3 &=&0~, \label{EoM1.a} \\
-\frac 14 \bar \nabla^2 \big(\bar \F \F^{-2} \big) &=& \m~.\label{EoM1.b}
\eea
The equation of motion corresponding to the gravitational superfield proves to be
\bea
\big[ \nabla_\a , \bar \nabla_\ad \big] \big( \bar \F \F\big)^{-1/2} =0~. 
\label{7.2c}
\eea
\end{subequations} 
In general, given a primary real scalar $L$ of dimension $-1$, ${\mathbb D} L = -L$, 
its real vector descendant $\big[ \nabla_\a , \bar \nabla_\ad \big] L$ is primary.

The equations \eqref{EoM1} can be degauged to $\sU(1)$ superspace, which results in 
\begin{subequations}\label{EoM2}
\bea
\bar \cD_\ad \F^3 &=&0~, \\
-\frac 14 \big(\bar \cD^2 -4R\big) \big(\bar \F \F^{-2} \big) &=& \m~, \label{EoM2.b}\\
\Big\{ G_{\a\ad} + \big[ \cD_\a , \bar \cD_\ad \big] \Big\}\big( \bar \F \F\big)^{-1/2} &=&0~. 
\label{EoM2.c}
\eea
\end{subequations} 
Now, the super-Weyl and local $\sU(1)_R$ gauge freedom can be used to impose the gauge condition $\F=1$. This implies that the $\sU(1)_R$ connection vanishes, and $\sU(1)$ superspace geometry reduces to the GWZ geometry. The supergravity equations \eqref{EoM2.b} and \eqref{EoM2.c} turn into 
\bea
R= \m = {\rm const}~, \qquad G_{\a\ad} =0~,
\label{EoM7.4}
\eea
and all information about the dynamics of supergravity is encoded in the super-Weyl tensor 
$W_{\a\b\g}$. By analogy with the terminology used in general relativity, the equations \eqref{EoM7.4} define an Einstein superspace. 

A unique maximally supersymmetric solution of \eqref{EoM7.4} 
 is characterised by the condition $W_{\a\b\g} =0$. It is called $\cN=1$ AdS superspace, and can be identified with the homogeneous space 
$ {{\sOSp}(1|4)}/{{\sSO}(3,1) }$ 
introduced in 
\cite{Keck,Zumino77}. The superfield representations of AdS supersymmetry were classified by Ivanov and Sorin \cite{IS}.


\subsection{New minimal  supergravity}

In the new minimal formulation for Poincar\'e supergravity, the compensator 
is a tensor multiplet  $\mathbb G$ obeying the constraints \eqref{6.22}.
The supergravity action is given by the functional 
\bea
S_{\text{new-minimal}}=3\int {\rm d}^{4|4}z \,E\, \mathbb G \ln \frac{\mathbb G } {\bar \f \f}
~,
\label{TypeII_action}
\eea
which differs only by a negative overall factor from \eqref{ImprovedTensor}.
The equation of motion for the chiral spinor prepotential $\eta_\a$, eq. \eqref{6.19}, is 
\bea
\bar \nabla^2 \nabla_\a  \ln \frac{\mathbb G } {\bar \f \f} = 0~,
\eea
and its general solution is given by 
\bea
{\mathbb G } = \bar \F \F~, \qquad 
K^B \F =0~, \quad  {\bar \de}^\bd \F=0~, \quad
  {\mathbb D} \F = \F~. 
  \label{7.7} 
  \eea
  Here the chiral scalar $\F$ is nowhere vanishing. 
  Now, the constraint $\bar \nabla^2 \mathbb G =0$ leads to the equation on $\bar \F$ 
  \bea
  \bar \nabla^2 \bar \F =0~,
  \eea
  which is equivalent to the equation \eqref{EoM1.b} with $\m =0$.
  The equation of motion for the gravitational superfield can be shown to be equivalent to  
 \bea
 \nabla_\a  \ln {\mathbb G} \bar \nabla_\ad \ln {\mathbb G}
  - \big[\nabla_\a, \bar \nabla_\ad \big]\ln {\mathbb G} =0~,
 \eea
 where the left-hand side proves to be a primary real vector superfield. 
This equation may be seen to be equivalent to \eqref{7.2c} if the representation \eqref{7.7} for 
$ {\mathbb G}$ in terms of $\F$ is used. 

The above results imply that new minimal supergravity is classically 
equivalent to old minimal supergravity without cosmological term.


\subsection{Three-form supergravity} 

The only difference of three-form supergravity from the old minimal theory considered earlier is that the chiral compensator $\F$ is realised in the former theory as $\F = \P^{1/3}$, where $\P$ is the chiral field strength of the three-form multiplet, eq. \eqref{6.24}.
The supergravity action is 
\bea
S_{\text{three-form}} &=& \int\text{d}^{4|4}z \, E \, \Big\{ -3\big(\bar \P \P\big)^{1/3} +  2m P \Big\} 
~,
\label{three-form_model}
\eea
with $m$ a real coupling constant. The equation of motion for $P$ takes the simplest form in terms of $\F = \P^{1/3}$ and its conjugate: 
\bea
-\frac 14 \bar \nabla^2 \big( \bar \F \F^{-2} \big) -\frac 14  \nabla^2 \big( \F \bar \F^{-2} \big) 
= 2m~.
\eea
This is equivalent to the equation
\bea
-\frac 14 \bar \nabla^2 \big( \bar \F \F^{-2} \big) = \m = {\rm const}~, \qquad 
{\rm Re} \,\m = m
\eea
which has the same form as \eqref{EoM1.b}. The new feature of the supergravity theory 
\eqref{three-form_model} is that the imaginary part of $\m$ is now generated dynamically.
It may be shown that the equation of motion for the gravitational superfield is equivalent to 
\eqref{EoM2.c}. As a result, the supergravity theories \eqref{old-minimal-action}
and \eqref{three-form_model} are classically equivalent. 

The existence of three-form supergravity was first pointed out by Gates and Siegel
\cite{GS81}. Unlike the standard formulation of old minimal supergravity, 
the remarkable feature of three-form supergravity is that it 
allows a consistent coupling to the 
four-dimensional  supermembrane \cite{Achucarro:1988qb}
as demonstrated by Ovrut and Waldram \cite{OvrutWaldram} who built on the results of 
 \cite{Binetruy:1996xw}. Within the GWZ formalism, the action \eqref{three-form_model} was presented in \cite{KMcC}.


\subsection{Non-minimal supergravity}

In the  non-minimal formulation for Poincar\'e supergravity,
the compensator is a complex linear superfield $\G$ constrained by \eqref{complex-linear}.
The supergravity action is described by the functional  
\bea
S_{\text{non-minimal}} = \frac{1}{n} \int \rd^{4|4}z\,  E\, \big( \bar\G \G\big)^{ (3n+1) /2 }
~,
\eea
which differs from \eqref{non-min6.29} by an overall sign. It may be shown that  non-minimal supergravity is classically 
equivalent to old minimal supergravity without cosmological term, see \cite{GGRS,BK} for reviews. No supersymmetric cosmological term is allowed in non-minimal supergravity with the compensator $\G$ \cite{GGRS}.

To describe non-minimal AdS supergravity \cite{BK11}, the compensator $\U$ is chosen to obey the constraints \eqref{AdScompensator}, and the action is
\bea
S_{\text{non-minimal AdS}} &=& - \int \rd^{4|4}z\,  E\, \big( \bar\U \U\big)^{-1}~.
\label{S-nm-AdS}
\eea
In order to derive the equation of motion for $\U$, we note that $\d \U$ is a complex linear superfield and hence $\d \U = \bar \nabla_\ad \d \bar \J^\ad$. Varying the action gives 
\bea
\bar \nabla_\ad \big( \U^2 \bar \U\big)^{-1} =0\quad \implies \quad \big(\U^2 \bar \U\big)^{-1} = \F^{3}
~.
\eea
We see that the equation of motion for $\U$ in the non-minimal theory \eqref{S-nm-AdS}
is equivalent to the off-shell constraint  \eqref{EoM1.a} in old minimal supergravity. 
It follows that $\U = \bar \F \F^{-2} $, and the off-shell constraint \eqref{AdScompensator} turns into the equation of motion \eqref{EoM1.b} in old minimal supergravity. Finally, it may be shown that, in the non-minimal theory \eqref{S-nm-AdS},
the equation of motion for the gravitational superfield  is equivalent to \eqref{7.2c} once $\U$ is expressed in terms of $\F $ and its conjugate.
We conclude that the minimal and non-minimal formulations for AdS supergravity, 
which are described by the actions \eqref{old-minimal-action} and \eqref{S-nm-AdS},
 are classically equivalent. 


\subsection{Conformal supergravity}

The conformal supergravity action  \cite{FZ2,Siegel78,ZuminoSS}, is 
\bea
S_{\rm CSG}= -\frac{1}{4} \int \rd^4x 
\rd^2 \q \, \mathcal{E} \, {W}^{\a \b\g } W_{\a\b\g}+ \text{c.c.}
\eea
The corresponding equation of motion is 
\bea
B_{\a\ad} = \bar B_{\a\ad}=0~,
\eea
with $B_{\a\ad}$ being the super-Bach tensor \eqref{super-Bach}.
This equation can be degauged to $\sU(1)$ superspace to take the form
\begin{align}
 \ri \cD_{\b \ad} \cD_\g W^{\a\b\g}
+  \cD_\b (G_{\g \ad} W^{\a\b\g})
= \ri \cD_{\a \bd} \bar \cD_\gd \bar W^{\ad\bd\gd}
- \bar \cD_\bd (G_{\a \gd} \bar W^{\ad\bd\gd})
=0~.
\end{align}
It follows from the Bianchi identities \eqref{Bianchi1} and  \eqref{4.10c} 
that every solution of the equations of motion for pure AdS supergravity  \eqref{EoM7.4} is also a solution of the equations of motion for conformal supergravity. 
\\

\noindent
{\bf Acknowledgements:}\\ 
We thank S. James Gates Jr. for his kind invitation to contribute this chapter to the {\it Handbook of Quantum Gravity}.
We are grateful to Daniel Butter for useful discussions, and to Stefan Theisen for comments on the manuscript.
The work of SK  is supported in part by the Australian 
Research Council, project No. DP200101944.
The work of ER is supported by the Hackett Postgraduate Scholarship UWA,
under the Australian Government Research Training Program. 
The work of GT-M is supported by the Australian Research Council (ARC)
Future Fellowship FT180100353, and by the Capacity Building Package of the University of Queensland.


\end{document}